\documentclass[letter,12pt]{article}

\usepackage[T1]{fontenc}
\usepackage{amsmath}
\usepackage{amsfonts}
\usepackage{amssymb}
\usepackage{color}
\usepackage{ upgreek }
\usepackage{jheppub}

\newcommand{\bea}{\begin{eqnarray}}
\newcommand{\eea}{\end{eqnarray}}

\newcommand{\CC}{\ensuremath{\mathcal C}}

\newcommand{\g}{ g}
\newcommand{\genus}{{\sf g}}
\newcommand{\I}{\ensuremath{\mathbb I}}

\newcommand{\R}{\ensuremath{\mathbb R}}

\newcommand{\Z}{\ensuremath{\mathbb Z}}

\title{Holographic description of Narain CFTs and their code-based ensembles}
\author[a,b]{Ofer Aharony,}
\author[c]{Anatoly Dymarsky,} 
\author[c]{and Alfred D. Shapere} 
 
\affiliation[a]{Department of Particle Physics and Astrophysics, Weizmann Institute of Science, \\
Rehovot 7610001, Israel}
\affiliation[b]{School of Natural Sciences, Institute for Advanced Study, Princeton, NJ 08540, USA}
\affiliation[c]{Department of Physics and Astronomy,  University of Kentucky,\\ 506 Library Drive, Lexington, KY, USA 40506}

\emailAdd{Ofer.Aharony@weizmann.ac.il}
\emailAdd{a.dymarsky@uky.edu}
\emailAdd{shapere@g.uky.edu}

\abstract{
We provide a  precise relation between an ensemble of Narain conformal field theories (CFTs) with central charge $c=n$, and a sum of $(U(1) \times U(1))^n$ Chern-Simons theories on different handlebody topologies. We begin by reviewing the general relation of additive codes to Narain CFTs. Then we describe a holographic duality between any given  Narain theory and a pure  Chern-Simons theory on a handlebody manifold. 
We proceed to consider an ensemble of Narain theories, defined in terms of an ensemble of codes of length $n$ over $\Z_k \times \Z_k$ for prime $k$. We  show that averaging over this ensemble is holographically  dual to a level-$k$ $(U(1) \times U(1))^n$ Chern-Simons theory, summed over a finite number of inequivalent classes of handlebody topologies. 
In the limit of large $k$ the ensemble approaches the ensemble of all Narain theories, and its bulk dual becomes equivalent to ``U(1)-gravity'' -- 
the sum of the pertubative part of the Chern-Simons  wavefunction over all possible handlebodies -- providing a bulk microscopic definition for this theory. 
Finally, we reformulate the sum over handlebodies  in terms of Hecke operators,  paving the way for  generalizations. 
}

\begin{document} 
\maketitle
\flushbottom

\section{Introduction}
\label{sec:intro}
In recent years, it has become evident that certain gravitational theories in anti-de Sitter (AdS) space are  dual to ensemble averages, rather than to individual quantum field theories.  
A general argument for requiring an ensemble to describe gravitational bulk theories is based on the presence of bulk geometries with several disconnected boundaries \cite{Maldacena:2004rf}, known as ``wormholes''.   If the gravitational action of such configurations is non-trivial, 
then the dual field theory will not factorize on disconnected manifolds, necessitating
an ensemble interpretation.  The first explicit example of such a  duality arose in 2d JT gravity, which was found to be dual  to an average over an ensemble of quantum-mechanical systems \cite{Saad:2019lba}. In one dimension higher, an intriguing  example that motivated 
our study is provided  by a theory called ``U(1) gravity'', which is formulated as a sum over handlebody geometries in the bulk, and is dual to an average over the moduli space of Narain CFTs \cite{Afkhami-Jeddi:2020ezh,Maloney:2020nni} (see also 
\cite{perez2020gravitational,Dymarsky:2020pzc,Datta:2021ftn,Benjamin:2021wzr,Benjamin:2021ygh,Meruliya:2021lul,Dong:2021wot,Ashwinkumar:2021kav,Collier:2021rsn,Chakraborty:2021gzh,Raeymaekers:2021ypf,Benini:2022hzx,kamesking2023lion}
 for further developments). Yet the original example of a holographic correspondence, between the ${\mathcal  N}=4$ supersymmetric Yang-Mills theory and type IIB string theory on $AdS_5\times S^5$ \cite{Maldacena:1997re}, has so far evaded an ensemble interpretation.  This raises the question: when does an ensemble of field theories admit a holographic interpretation?  In particular, can a finite ensemble have a gravitational dual, and  which bulk geometries need to be summed over in this case? 

In this paper, we address the latter question by studying  finite ensembles of Narain theories  composed of ``code CFTs'', which were introduced in \cite{Dymarsky:2020qom,Yahagi:2022idq,Angelinos:2022umf}. We find that they are dual to a $(U(1)\times U(1))^n$ Chern-Simons theory of finite level, summed over a finite number of inequivalent handlebody topologies.  

The relation between error-correcting codes and CFTs goes back all the way to the Golay code, which is associated with the Leech lattice, and which led to the discovery of the Monster CFT  \cite{frenkel1989vertex}, followed by other developments connecting codes and chiral CFTs \cite{dolan1996conformal,miyamoto1996binary,dong1998framed,LAM2000268,gaiotto_johnson-freyd_2022,Moriwaki:2021ebe}. 
Motivated by these developments, as well as by the emergence of quantum error correction in the context of  bulk reconstruction \cite{Almheiri:2014lwa},  two of us proposed a connection between quantum codes and non-chiral CFTs in \cite{Dymarsky:2020qom}. 
Our work led to further activity connecting codes and CFTs, with applications to the modular bootstrap program and beyond \cite{Dymarsky:2020bps,Dymarsky:2020pzc,Dymarsky:2021xfc,Buican:2021uyp,  Angelinos:2022umf,Yahagi:2022idq,Henriksson:2021qkt,Henriksson:2022dml,Henriksson:2022dnu,Dymarsky:2022kwb,Kawabata:2022jxt,furuta2023rationality,alam2023narain,kawabata2023supersymmetric,Kawabata:2023nlt,kawabata2023elliptic,Kawabata:2023iss}. 
Ensembles of code CFTs were found in  \cite{Dymarsky:2020pzc, Angelinos:2022umf} to be self-averaging and to exhibit a large spectral gap, suggesting a possible holographic interpretation and motivating the current study. 

We first consider the holographic description of an individual  Narain CFT with $c=n$ on a Riemann surface $\Sigma$.  By explicitly evaluating the partition functions on both sides of the duality for $\Sigma$ of genus one, we show that it is dual to a pure level-1 $(U(1) \times U(1))^n$ ``$AB$''  Chern-Simons (CS) theory  on a 3-manifold $\mathcal M$ with  boundary $\partial {\mathcal M}=\Sigma$ (any such 3-manifold can be chosen and gives the same results, there is no sum over 3-manifolds), and we establish the precise holographic dictionary.  
We note that the two $U(1)^n$ gauge fields are coupled at the level of large gauge transformations, and their boundary conditions determine the moduli of the Narain theory.  
The level $k=1$ Chern-Simons theory avoids the factorization puzzle  because it is trivial in the bulk -- it has a unique wavefunction on any $\Sigma$, and in particular the partition function on a ``wormhole'' geometry  connecting two disjoint boundaries is the same as that on the disconnected product of manifolds with the same boundaries.  

For a $(U(1) \times U(1))^n$ CS theory  of level $k>1$, the field theory dual is no longer an individual Narain CFT.  Rather, we find it to be dual to an ensemble average over a finite set of $c=n$ Narain CFTs based on the set of all even self-dual codes of length $n$ over  the ``alphabet'' $\Z_k \times \Z_k$.  In this case the CS wavefunction depends on the topology of the  bulk manifold $\mathcal M$, and we find that the averaged CFT partition function is precisely reproduced by the corresponding Chern-Simons wavefunction, summed over a finite number of equivalence classes of handlebody topologies. The boundary conditions of the CS theory map to parameters of the Narain theories in the ensemble. Our main identity  \eqref{ensembleH}, valid for any fixed $n$ and prime $k=p$, gives an explicit relation between the average over the code-based ensemble and the ``Poincar\'e series'' representing a (finite) sum over bulk geometries.  The $k=1$ duality of the previous paragraph may be viewed as a special case of this, where the ensemble contains just a single CFT.

As  $p\to \infty$, for $n>2$ we will argue that the ensemble of code theories densely covers the whole of Narain moduli space with the canonical measure.  We show explicitly how a similar limit works in the case of $n=2$, by expressing the average in terms of Hecke operators, and applying a theorem  \cite{clozel2001hecke} on the equidistribution of Hecke points.  In the bulk, for $n>2$ in the  $p\to \infty$ limit we recover the full Poincar\'e sum over all handlebody topologies,  reproducing the ``U(1)-gravity'' of \cite{Afkhami-Jeddi:2020ezh,Maloney:2020nni}. Thus, our construction provides a microscopic bulk definition for the latter, as a limit of CS theories.  

Arguments of typicality suggest that for large ensembles of CFTs that are self-averaging and possess a holographic description as a sum over geometries, random individual theories should also admit an approximate holographic description as a sum over geometries. Motivated by this, we propose a sum-over-geometries description for any individual Narain theory with $n >2$, that in general is non-local in the bulk, but that becomes approximately local for typical (random) theories as $n\to \infty$ (which is the limit in which the ensemble becomes self-averaging).

The plan of the paper is as follows. 
In Section \ref{sec:codes}, we briefly review the relation between additive codes, lattices, and Narain CFTs.  In the course of this discussion, we generalize previous constructions by introducing arbitrary additive codes in Section \ref{sec:gencodes}.  Section \ref{sec:CS} reviews $(U(1)\times U(1))^n$ Chern-Simons theories on handlebody geometries, and constructs their wavefunctions for general boundary conditions.   In Section \ref{sec:hol} we discuss the holographic interpretation of Narain theories and their ensembles. First,  in section \ref{sec:convhol} we show that the wavefunction of level-1 $(U(1)\times U(1))^n$ Chern-Simons theory, evaluated with given boundary conditions, is equal to the partition function of a Narain CFT.  The point in the Narain moduli space is specified by the  boundary conditions of the CS theory, establishing an explicit holographic dictionary.  
We briefly discuss the idea of averaging over these boundary conditions in section \ref{sec:bcaveraging}, and proceed to discuss level $k>1$  CS theories summed over geometries in section \ref{sec:Pseries}. This section establishes our main technical result, equation (\ref{ensembleH}).  We discuss the $k\to \infty$ limit and the emergence of ``U(1)-gravity'' in section \ref{sec:kinf}. Section \ref{c=1c=2} is devoted to a detailed analysis of the $n=1$ and $n=2$ cases. It also establishes connections with Hecke operators, which we further discuss, together with related mathematical observations, in section \ref{sec:ext}. 
Section \ref{sec:typicality} explores the holographic description of an individual Narain theory as a sum over geometries. We conclude with a discussion in Section \ref{sec:Discussion}. Several appendices contain technical details.

\section{Additive codes and Narain CFTs}
\label{sec:codes}
\subsection{Codes over $\Z_k\times \Z_k$}
A classical additive code over an Abelian group $F$ is a collection of  $F$-valued strings (codewords) of length $n$ closed under addition within $F$.  Additive codes are naturally related to lattices \cite{conway2013sphere}, and thus to lattice-based chiral CFTs \cite{dolan1996conformal}.  Recently, codes of more general type have been shown to be  related to Narain CFTs, their orbifolds, and Abelian fermionic theories \cite{Dymarsky:2020qom,Dymarsky:2020bps,Dymarsky:2021xfc,Angelinos:2022umf,Yahagi:2022idq,Henriksson:2022dnu,Kawabata:2022jxt,furuta2023rationality,alam2023narain,kawabata2023supersymmetric,Kawabata:2023nlt,kawabata2023elliptic,Kawabata:2023iss}.

As an illustrative example, we briefly review the relation between additive even self-dual codes over $\Z_k \times \Z_k$ and Narain theories \cite{Yahagi:2022idq,Angelinos:2022umf}. A code ${\mathcal C}$ over $\Z_k \times \Z_k$ can be thought of as a linear space within $\Z_k^{2n}$ where all algebra is defined mod $k$. The space is equipped with the indefinite scalar product 
\bea
\label{eta}
\eta=\left(\begin{array}{cc}
0 & \I_{n\times n}\\
\I_{n\times n} & 0
\end{array}\right)\label{sp2n}
\eea 
with respect to which all code vectors are ``even'', $c^T \eta\, c=0\, ({\rm mod}\ 2k)$ for all $c\in \mathcal C$.
Furthermore, self-duality implies that any vector $c'\in \Z^{2n}$ orthogonal to $\mathcal C$, in the sense that $c^T \eta\, c'=0\, ({\rm mod}\ k)$ for any $c\in \mathcal C$, also belongs to $\CC$.  There are 
\bea
{\mathcal N}=\prod_{i=0}^{n-1}(p^i+1) \label{numberofcodes}
\eea
distinct codes of this type when $k=p$ is prime (the expression for composite $k$ is more involved).  
Starting from an even self-dual code $\CC$, we can define an even self-dual lattice $\Lambda_{\mathcal C} \subset \R^{n,n}$ as follows:
\bea
\label{LC}
\Lambda_\CC\equiv\left \{ v/\sqrt{k}\, |\,  v\in \Z^{2n}, \, v\, {\rm mod}\, k\in \CC \subset \Z^{2n}_k \right\}.
\eea
A lattice $\Lambda_\CC$ defines a Narain CFT. When $k=p$ is prime, the CFT can be described,  via a $T$-duality transformation\footnote{T-duality is commonly understood as the action of $O(n,n,\Z)$ on $\gamma$ and $B$. From the lattice generator matrix point of view, it is the action of $O(n,n,\Z)$ from the right, amended by the action of $O(n,\R)\times O(n,\R)$ from the left to preserve the ``left bottom block equal zero'' structure as in \eqref{LCC}. From the Narain lattice point of view action of $O(n,n,\Z)$ from the right is trivial. Hence, in the context of Narain lattices, by  T-duality we mean the action of $O(n,\R)\times O(n,\R)$.}, as a compactification on an $n$-dimensional torus with metric $\gamma=I/\sqrt{p}$ and with $B$-field given by an antisymmetric integer-valued matrix ${\rm B}_{ij}\in \Z$,  such that $G=(I,{\rm B}^T)$ with ${\rm B}=B\, {\rm mod}\, p$ is the generator matrix of the code $\CC$ brought into canonical form,\footnote{The generator matrix $G$ is a $n\times 2n$ matrix such that all codewords are given by $c=G^T q\, {\rm mod}\, k$, $q\in \Z_k^n$. The form of $G=(I,{\rm B}^T)$ with some antisymmetric ${\rm B}_{ij}\in \Z_k$ is called canonical. When $k$ is prime, one can always bring the generator matrix to the canonical form using so-called code equivalence transformations 
\cite{Angelinos:2022umf}.}
\bea 
\label{LCC}
\Lambda_\CC  = O_T \left(\begin{array}{cc}
\gamma^* & \gamma^* B\\
0 & \gamma\end{array}\right),\qquad \gamma^*\equiv (\gamma^{\mathsf{T}})^{-1},  \qquad O_T\in O(n,\R)\times O(n,\R).
\eea
Here we use $\Lambda_\CC$ to denote both the lattice and the lattice-generating matrix. 

An important object characterizing a code is the complete enumerator polynomial $W_\CC$. It counts the number of codewords of a code, that include a given ``letter'' with the specified multiplicity. In the present case, with the ``alphabet'' $\Z_k \times \Z_k$, we regard a codeword $c=(a_1,\dots, a_n,b_1,\dots, b_n)$ as being composed of letters $(a_i,b_i)\in \Z_k\times \Z_k$.  Introducing $k^2$ formal variables $X_{ab}$  with $0\leq a,b<k$ to represent the letters, one defines the complete  enumerator polynomial
\bea
W_\CC(X)=\sum_{(\vec{a},\vec{b})\in \CC} \prod_{i=1}^n X_{a_i b_i}. \label{WC}
\eea 
For self-dual $\CC$, 
$W_\CC$ satisfies the so-called Mac-Williams identity 
\bea
\label{MW}
W_\CC(X)=W_\CC(X'),\quad {\rm where\ \ }X'_{ab}\equiv{1\over k} \sum_{a',b'} X_{a' b'}e^{-2\pi i (a'b+a\, b')/k}.
\eea

To better illustrate the notation, we consider a simple example -- length $n=1$ even self-dual codes over $\Z_k\times \Z_k$. When $k=1$ there is a unique code consisting of only one codeword $(0,0)\in\CC$. When $k=p$ is prime, there are two codes, one with codewords of the form $(a,0)\in \CC_1$ and the other with $(0,b)\in \CC_2$, with arbitrary $0\leq a,b<p$. Their enumerator polynomials are 
\bea
W_{\CC_1}(X)=\sum_{a=0}^{p-1} X_{a0},\qquad W_{\CC_2}(X)=\sum_{b=0}^{p-1} X_{0b}.
\eea
When $k>1$ is not prime, there are more codes. All length $n=2$ codes for prime $k=p$ are listed in Appendix \ref{sec:n=2codes}.

A defining feature of the code construction of Narain theories is that for a Narain theory defined with the help of a code-based lattice \eqref{LC}, its torus partition function can be concisely written in terms of $W_\CC$. Indeed, the torus partition function of a Narain theory is defined in terms of a Siegel theta series that sums over all lattice points. For $\Lambda_\CC$ as in \eqref{LCC}, we can readily see that the lattice points organize into sets, each associated with a given codeword $(\vec a,\vec b)\in \CC$:
\bea
S_{\vec a,\vec b}=\{ v/\sqrt{k}\in \Lambda_\CC\,|\,v = (\vec a,\vec b)\,\,  {\rm mod}\, \, k \}. 
\eea
We can sum over these sets  individually, yielding 
\bea
Z(\tau)&=&W_\CC(\Psi),\quad \Psi_{ab}={\Theta_{ab}\over |\eta(\tau)|^2}, \nonumber\\ \Theta_{ab}&\equiv&\sum\limits_{n,m} e^{i\pi \tau p_L^2-i \pi \bar \tau p_R^2},\quad p_{L,R}=\sqrt{k\over 2}((n+a/k)\pm (m+b/k)), \quad n,m\in\Z. \label{psiabzkzk}
\eea
It can be readily seen that by virtue of $\CC$ being even, each combination $\prod_{i=1}^n \Psi_{a_ib_i}$ in $W_\CC$ associated with an individual codeword $(\vec{a},\vec{b})\in \CC$ will be invariant under $\tau\rightarrow \tau+1$, although individual factors
\bea
\label{Ev}
\Psi_{ab}(\tau+1)=\Psi_{ab}(\tau)\, e^{2\pi i a b/k}
\eea 
are not. 
  Furthermore, $Z$ will be invariant under $\tau \rightarrow -1/\tau$ due to the Mac-Williams identity \eqref{MW} and the fact that   $\Psi_{ab}(-1/\tau)=\Psi'_{ab}(\tau)$,
where  $\Psi'$ are defined as $X'$ in \eqref{MW}.
 The  relation between the code's enumerator polynomial and the associated CFT partition function can be extended to higher genus \cite{Henriksson:2021qkt,Henriksson:2022dnu}.

The relation between codes and CFTs at the level of the partition function has proved to be a useful tool, which among other things provides an efficient way to solve modular bootstrap constraints, construct inequivalent isospectral CFTs \cite{Dymarsky:2020bps} and modular invariant  $Z(\tau)$ which are ``fake'' (i.e.,~not associated with any bona fide CFT) \cite{Dymarsky:2022kwb}, construct ``optimal'' CFTs maximizing the value of the spectral gap \cite{Dymarsky:2021xfc,Angelinos:2022umf}, etc. One recent application was the calculation of the spectral gap of $U(1)^n \times U(1)^n$ primaries for a typical code theory when $k\rightarrow \infty$  while $n$ is kept fixed \cite{Angelinos:2022umf}. The resulting gap, $\Delta=n/(2\pi e)$, matches the value of the spectral gap in ``$U(1)$-gravity'' \cite{Dymarsky:2020pzc,Afkhami-Jeddi:2020ezh}, a result we return to in section \ref{sec:typicality}.

The results mentioned above mostly rely on a ``rigid embedding'' \eqref{LC} or its analogues, in which a code, understood as a subset of $\Z_k^{2n}$, is mapped to a lattice $\Lambda_\CC $, which is a sublattice  of a cubic lattice of spacing $1/\sqrt{k}$, 
$(\sqrt{k}\, \Z)^{2n}\subset \Lambda_\CC \subset (\Z/\sqrt{k})^{2n}\subset \R^{n,n}$.  This rigidity, which allows only very special Narain lattices to be obtained from codes, suggests a picture in which codes are related to a set of very special Narain theories, dubbed code CFTs. In this picture, there is a close relation between the underlying code and the algebraic properties of the CFT \cite{Buican:2021uyp}. However, as we will see momentarily, these maps from codes to CFTs are a particular instance of a much more general relation. 

\subsection{General case}
\label{sec:gencodes}
Reference  \cite{Angelinos:2022umf} provides a general construction of codes over an Abelian group ${\sf G}$ defined as the quotient group of a self-orthogonal even ``glue lattice'' 
${\sf\Lambda}$, 
\bea
{\sf G}={\sf\Lambda}^*/{\sf\Lambda},\qquad {\sf\Lambda}^*\equiv \eta ({\sf\Lambda}^\mathsf{T})^{-1}. 
\eea
In \cite{Angelinos:2022umf} the focus was on ${\sf\Lambda}\subset  \R^{1,1}$ and all such lattices were classified there. They are defined by
\bea
\label{Ldef}
{\sf\Lambda}^T \eta {\sf\Lambda}=g_{\sf\Lambda}=\left(\begin{array}{cc}
2\, {\sf n} & {\sf k}\\
{\sf k} & 2\, {\sf m}\end{array}\right), \quad {\sf n}, {\sf m},{\sf  k}\in \Z, \quad {\sf k}^2-4{\sf nm}>0,
\eea
with an arbitrary $O_+(1,1)$ transformation acting on ${\sf\Lambda}$. In particular, the case of $\Z_k \times \Z_k$ codes discussed above corresponds to the glue matrix
\bea
{\sf\Lambda}=\left(\begin{array}{cc}
1/r & 0 \\
0 & r\end{array}\right)\sqrt{k} \label{Lambdak}
\eea
with $r=1$.  A nontrivial  ``embedding'' $r$ in \eqref{Lambdak} changes $p_{L,R}$ in \eqref{psiabzkzk} to
\bea
p_{L,R}=\sqrt{k\over 2}((n+a/k)/r\pm (m+b/k)r),
\eea
while changing neither the relation between $Z(\tau)$ and $W_\CC$, nor the way in which $\Psi_{ab}(\tau)$ changes under modular transformations of $\tau$. The group $\sf G$ is an ``alphabet'', while codes are collections of $\sf G$-valued strings of length $n$ closed under addition and equipped with the scalar product inherited from $\eta$. Then, even self-dual codes $\CC$ over $\sf G$ define even self-dual (Narain) lattices $\Lambda_\CC$ in $\R^{n,n}$ via a straightforward generalization of \eqref{LC},
\bea
{\Lambda} \subset\,   \Lambda_\CC\, \subset {\Lambda}^* \subset \R^{n,n},\qquad {\Lambda}= {\sf\Lambda}\oplus \dots \oplus {\sf\Lambda}. \label{glue}
\eea
If all $n$ glue lattices ${\sf\Lambda}$ in \eqref{glue} are the same, then the permutation of letters within the codeword -- a code equivalence -- is also a symmetry (an element of the T-duality group) at the level of the code CFT. But one can also  choose $n$ different parameters $r_i$, in which case to preserve the relation between $Z$ and $W_\CC$, the enumerator polynomial should depend on $n\, k^2$ distinct auxiliary variables $X_{ab}^i$, $W_\CC=\sum_{(\vec{a},\vec{b})\in \CC}\prod_i X^i_{a_i b_i}$. More generally, one can consider $O(n,n,\R)$ transformations acting and mixing several or all ${\sf\Lambda}$'s within $\Lambda$, or combinations of completely different even self-orthogonal matrices ${\sf\Lambda}$ of different dimensions (leading to codes where different letters belong to different alphabets). 

Thus, most generally one can consider a lattice ${\Lambda}\subset \R^{n,n}$, even and self-orthogonal with respect to the $2n$-dimensional scalar product \eqref{sp2n}
within $\R^{n,n}$, with the ``codewords'' being elements of the Abelian quotient group $c\in \CC \subset{\Lambda}^*/{\Lambda}={\sf G}_{\Lambda}$. This group defines a ``dictionary,'' a set of all possible ``words''. The ``dictionary group''  inherits the scalar product from \eqref{sp2n}. An even self-dual code would additionally satisfy 
\bea
&&c^T\, \eta\, c \in 2\Z\, \, \,  {\rm for\, \, any}\, \, c\in \CC,\\
&&c^T \, \eta\, c' \in \Z\, \, \,  {\rm for\, \, all}\, \, c,c'\in \CC, 
\eea
while if $c'\notin \CC$ then $c^T \, \eta\, c'$ is not an integer for some $c\in \CC$. 

Any even self-dual code then defines a Narain lattice, generalizing \eqref{LC},
\bea
\Lambda_\CC=\left\{v\, |\,  v\in {\Lambda}^*,\, (v\ {\rm mod}\,{\Lambda})\in \CC\right\}, \quad {\Lambda}\subset \Lambda_\CC \subset {\Lambda}^*\subset \R^{n,n}.
\eea 
Here we denote by $(v\ {\rm mod}\,{\Lambda})$ the equivalence class of $v$ within ${\Lambda}^*/{\Lambda}$.
In general, the relation between the associated CFT partition function and the code enumerator polynomial remains essentially the same. The complete enumerator polynomial  
\bea
W_\CC(X)=\sum_{c\in \CC} X_c,
\eea
is defined in term of formal auxiliary variables $X_c$ for $c\in {\sf G}_{\Lambda}$, which are then promoted to ``(code)word blocks'' $\Psi_c$ with modular parameter $\tau$ and  arbitrary fugacities $\xi, {\bar \xi}\in \R^n$
\bea
\label{Psic}
&&\Psi_c(\tau,\xi,\bar \xi)={{\Theta}_c\over |\eta(\tau)|^{2n}},\quad \Theta_c=\sum\limits_{\ell} e^{i\pi \tau p_L^2-i \pi \bar \tau p_R^2+2\pi i (p_L\cdot \xi-p_R\cdot \bar \xi)+{\pi\over 2\tau_2}(\xi^2+{\bar \xi}^2)}, \nonumber \\
&&\left(
\begin{array}{c}
p_L+p_R\\
p_L-p_R\\
\end{array}\right)=\sqrt{2}({{\Lambda}\, \vec{\ell}+\vec{c}}),\quad \vec{\ell}\in \Z^{2n},\quad {\vec c}\in \CC\subset {\sf G}_{\Lambda}\equiv {\Lambda}^*/{\Lambda} \label{PLR2}. 
\eea
We emphasize that $\Psi_c$ are defined for all $c\in{\sf G}_{\Lambda}$, and not all of them are even. 
The  {\it path integral} of the CFT  is then given by
\bea
\label{ZW}
&&Z_{BPI}=W_\CC(\Psi),
\eea
where ``BPI'' stands for bulk path integral. 
This name is justified in Section \ref{sec:CS}. 
$Z_{BPI}$  is equal to the CFT partition function up to a theory-independent factor, as explained in  Appendix \ref{App:Z}; see also \cite{Kraus_2007}.

The functions $\Psi_c$ change covariantly under  modular transformations
\bea
\Psi_c(\tau+1,\xi, \bar \xi)&=&\Psi_c(\tau,\xi, \bar \xi) e^{\pi i c^T \eta c}, \nonumber\\
\Psi_c(-1/\tau,\xi/\tau,\bar \xi/\bar\tau)&=&{1\over |{\sf G}_{\Lambda}|^{1/2}}\sum\limits_{c' \in {\sf G}_{\Lambda}} \Psi_{c'}(\tau,\xi, \bar \xi) e^{-2\pi i c^T \eta c'}.\label{ST} 
\eea
Modular invariance of $Z_{\it PI}$ follows from (\ref{ST}) and the algebraic properties of $W_\CC$ due to the evenness and self-duality (Mac-Williams identity) of the underlying code. The transformations (\ref{ST}) are defined solely in terms of the code  and therefore can be defined already at the level of the formal variables $X_c$. 

The same functions $\Psi_c$, with $\xi=\bar\xi=0$,  have been discussed in \cite{Ashwinkumar:2021kav}, where they appeared in a different context -- as the partition functions of non-modular-invariant CFTs. There, an ensemble of such CFTs generated by the action of $O(n,n,\R)$ on a given $\Lambda$  was discussed, together with its holographic interpretation. The focus in this paper is different: we sum over $\Psi_c$ for all $c$ belonging to a suitable even self-dual code such that the resulting partition function corresponds to a modular-invariant Narain CFT.

We would like to emphasize that the action of $O(n,n,\R)$ on $\Lambda$ does not affect the code, its enumerator polynomial, nor the transformation laws (\ref{ST}). It changes the ``embedding'' that maps the codes associated with a given $\Lambda$ into the space of Narain CFTs. Explicitly, this means we define $\Psi_c$ exactly as in \eqref{Psic}, but can introduce an arbitrary ${\mathcal O}\in O(n,n,\R)$,
\bea
&&\left(
\begin{array}{c}
p_L+p_R\\
p_L-p_R\\
\end{array}\right)={\mathcal O}\sqrt{2}({{\Lambda}\, \vec{\ell}+\vec{c}}),\quad \vec{\ell}\in \Z^{2n},\quad {\vec c}\in \CC\subset {\sf G}_{\Lambda}\equiv {\Lambda}^*/{\Lambda} 
\eea
where the change of notation can be absorbed into the definition of $\Lambda$. Choosing different ``embeddings'' $\mathcal O$ will change code theories, but the relation \eqref{ZW} between $Z$ and $W_\CC$ will remain the same.  Thus, starting from any ${\sf\Lambda}$, e.g.~as given by \eqref{glue} and \eqref{Lambdak} with $r=1$, and applying an appropriate $\mathcal O$, we can represent {\it any}  Narain lattice as a code lattice $\Lambda_\CC$  associated with {\it any}  $\CC$ over any alphabet. This, first of all, makes the notion that only certain Narain CFTs are associated with codes obsolete -- any Narain theory can be thought of as a code CFT associated with any even self-dual code of any type, i.e.~with any $\sf G$, $\CC\subset {\sf G}^n$, or more generally with any ${\sf G}_{\Lambda}\supset \CC$.  One can even associate several arbitrary Narain CFTs with several codes simultaneously, by making use of the $n(2n-1)$ parameters of $O(n,n,\R)$. 
Yet the notion of a code CFT ensemble is still relevant, since as $n$ increases there are generally many more codes of a given type, see e.g.~\eqref{numberofcodes}, than the number of adjustable parameters. 

In the case of $n=1$ codes with prime $k=p$ discussed above, the construction based on \eqref{Lambdak} gives two possible codes; the corresponding  CFTs are compact scalar theories with radii $R=r\sqrt{2p}$ and $R=r\sqrt{2/p}$, respectively. Obviously, by taking different values of $r$, each code covers the full space of $c=1$ Narain CFTs.

Another way to think about the relation of codes to CFTs  is that codes provide a simple tool to represent the modular invariant partition function $Z(\tau)$ of any given Narain theory as a sum of  ``codeword blocks'' $\Psi_c$ transforming in a particular representation of the modular group specified by $\sf G$ (or more generally ${\sf G}_{\Lambda}$) equipped with the scalar product. At a more technical level, the code -- a collection of individual codewords -- provides a division of a code-based Narain lattice into subsets $S_c$. The sum over such a subset is $\Psi_c$, which exhibits modular properties (\ref{ST}). Since all Narain lattices are related by the orthogonal group $O(n,n,\R)$, any code can be used to decompose any Narain lattice into subsets, such that the partial sums over the subsets will form a code-specified representation of the modular group.    

\section{$(U(1) \times U(1))^n$ Chern-Simons theories on a solid torus}
\label{sec:CS}

\subsection{A review of Abelian Chern-Simons theories on handlebodies}
\label{sec:gen_CS}
In this paper we discuss Abelian Chern-Simons theories on handlebodies, starting from a single Abelian factor, and then generalizing to many. Handlebodies are smooth 3-manifolds ${\mathcal M}$ whose boundary ${\partial {\mathcal M}}$ is a genus-$g$ Riemann surface $\Sigma$.   Topologically, they are characterized by the set of one-cycles of $\Sigma$ that are contractible in ${\mathcal M}$, which form a Lagrangian sublattice of $H^1({\mathcal M},\Z)$ with respect to the intersection form.
We will focus on the $g=1$ case, for which two important examples of handlebodies are thermal global anti-de Sitter space and (related to it by a modular transformation of the torus) the BTZ black hole. Since Chern-Simons theories are topological, the bulk metric of ${\mathcal M}$ will not play any role, but its topology and the metric on the boundary ${\partial {\mathcal M}}$ are important.

We begin with a $U(1)_k$ theory, whose Euclidean action is
\bea
\label{CSaction}
S={ik\over 4\pi} \int_{\mathcal M} A\wedge dA
\eea
for integer $k$. A natural way to study this theory on a handlebody is using radial quantization, where we view the handlebody ${\mathcal M}$ as a fibration of the Riemann surface $\Sigma$ over a ``radial'' direction (which can be viewed as Euclidean time) running from the interior (where the Riemann surface shrinks to zero volume) to the boundary. In the above examples of asymptotically $AdS_3$ spaces, the radial direction coincides with the usual radial coordinate of $AdS_3$. From this point of view, the quantum theory has a Hilbert space which is that of Chern-Simons theory on the Riemann surface, and the path integral over the handlebody evaluates  Euclidean time evolution starting from some initial state $|\Psi_{\rm interior}\rangle$ determined by boundary conditions in the interior, to some final state $|\Psi_{\rm boundary}\rangle$ corresponding to the boundary conditions on ${\partial {\mathcal M}}$. This Hilbert space contains $|k|^g$ states of zero energy, and any given state is a linear combination of these. The Euclidean time evolution is trivial, and the path integral on ${\mathcal M}$ is simply the overlap between the initial and final states inside this finite-dimensional Hilbert space, $\langle \Psi_{\rm boundary} | \Psi_{\rm interior} \rangle$. Note that for $|k|=1$ the Hilbert space is one-dimensional, so all handlebodies give rise to the same interior wavefunction.

In this radial quantization picture the two ``spatial'' components of $A$ along the Riemann surface are canonically conjugate variables, so they cannot both be diagonalized at the same time. One can choose to write the wavefunctions as functions of one or the other of these variables.  More precisely, we will express the wavefunctions in terms of holonomies of the gauge fields along $g$ nonintersecting one-cycles of the Riemann surface $\oint_{\gamma} A$; $W_{\gamma} = \exp(i \oint_{\gamma} A)$ is gauge-invariant, so the holonomies can be viewed as gauge-invariant coordinates up to shifts by $2\pi$ (which arise from large gauge transformations that preserve the wavefunction).
The holonomies for a basis of dual cycles of $\Sigma$ form a set of canonically conjugate variables. For any cycle $\gamma$ that shrinks in the interior of the handlebody, the interior wavefunction must obey $W_{\gamma} |\Psi_{\rm interior}\rangle= |\Psi_{\rm interior}\rangle$ (while the Wilson lines on the conjugate cycles are completely smeared).

In the presence of a boundary, having a consistent variational principle for the action \eqref{CSaction} requires that
\bea \label{variation}
\int_{\partial \mathcal{M}} A \wedge \delta A = 0
\eea
(involving both components of the gauge field along the boundary, and their variations). 
Equation \eqref{variation} can be satisfied by setting one of the components of the gauge field to zero at the boundary, or by setting to zero a complex combination of the two components, $A_z$ or $A_{\bar z}$, defined using an appropriate complex structure on $\partial \mathcal{M}$.  Setting a field to zero at the boundary automatically sets its variation to be zero.

In order to obtain more interesting possibilities for boundary conditions, one can add extra terms on the boundary. With an appropriate choice of a boundary term quadratic in $A$, one can cancel the terms in \eqref{variation} that involve a given component of $\delta A$ (either a spatial component or a complex combination), and then boundary conditions that set the other component of $A$ to any fixed value are also allowed. In particular, if (with an appropriate choice of boundary terms) $A_{\bar z}$ is frozen to a specific value at the boundary while $A_z$ on the boundary is allowed to fluctuate, then the Chern-Simons theory behaves as a chiral block of a 2d $U(1)_k$ CFT, with $A_{\bar z}$ interpreted as a source for a chiral $U(1)$ current $J(z)$ at level $k$ \cite{Witten:1988hf}. 

\subsection{The wavefunction of $U(1)_k$ theory on a torus}
\label{sec:CSU1}
As a warm-up example we construct the wavefunctions of level-$k$ $U(1)$ Chern-Simons  theory on a torus, following the classic work of Bos and Nair \cite{Bos:1989wa}. Additional technical details can be found 
in Appendix \ref{App:CS}.
We consider the CS theory \eqref{CSaction} on a three-dimensional manifold $\mathcal M$ with boundary $\partial \mathcal M$, which in our case will be a  torus with  modular parameter $\tau$.
We parametrize the boundary torus by the coordinate $z$, with identifications $z\sim z+1,z+\tau$. We choose a gauge where the radial component of the gauge field vanishes, 
and its equation of motion imposes $F_{z {\bar z}}=0$. 
We can then further choose a gauge where the spatial components of the gauge fields, $A_z$ and $A_{\bar z}$, are constant on the torus. 

Following \cite{Bos:1989wa, Elitzur:1989nr} we will consider a holomorphic representation of the wavefunction on the torus. 
This representation arises naturally if we deform the action \eqref{CSaction} by adding the boundary term 
 \bea \label{boundary_action}
S'=S-{k\over 2\pi}\int_{\partial M}d^2 z|A|^2,\qquad |A|^2 \equiv A_z A_{\bar z}, \qquad k>0,
 \eea
so that the equation of motion  $\delta S'/\delta A_z=0$  is trivially satisfied at the boundary. Then the path integral can be evaluated with boundary conditions of fixed $A_{\bar z}$ and arbitrary $A_z$.\footnote{Adding the term \eqref{boundary_action} with the opposite sign, which is natural for $k<0$, allows one to fix $A_z$ instead.}
The full path integral on the handlebody, including the boundary term, with a fixed value of $A_{\bar z}$
at the boundary (which is equivalent to the overlap with a wavefunction that is a delta function imposing the boundary value of $A_{\bar z}$) is then
\bea
\Psi_{\rm interior}(A_{\bar z})=\int\limits_{A_{\bar z}|_{\partial{\mathcal M}}\, \rm fixed} 
{\mathcal D}A\,  \, e^{-S'}. 
\eea
This is a holomorphic function of $A_{\bar z}$.

Because of the extra factor in the path integral, the overlap between two wavefunctions in the holomorphic representation is given by 
\bea \label{holo overlap}
\langle { \Psi_1} | { \Psi_2} \rangle = \int d^2A_{\bar z} \,(\Psi_1(A_{\bar z}))^* \Psi_2(A_{\bar z}) e^{-{k\over \pi} \int d^2z |A_{\bar z}|^2}.
\eea
This expression is schematic, since as we will discuss below one needs to remove degeneracies due to large gauge transformations; see Appendix \ref{App:CS} for details. 
The extra exponential factor in the overlap \eqref{holo overlap} can also be understood in the following way.
Understood as  quantum operators in radial quantization,  the gauge field components on the torus with  action \eqref{CSaction} obey the commutation relation $[A_z, A_{\bar z}] = \frac{\pi}{k\tau_2}$,\footnote{Note that this is not the same commutation relation that one obtains by starting at high energies in a Maxwell-Chern-Simons theory \cite{Gukov:2004id}, for which the phase space is labelled by $A_z$, $A_{\bar z}$ and their (independent) conjugate momenta. Here we describe wavefunctions on a different phase space, which is labeled only by $A_z$ and its canonical conjugate $A_{\bar z}$. See also \cite{Faddeev:1988qp,Jackiw:1993in}.}  so if we choose the wavefunctions to be functions of only $A_{\bar z}$, $A_z$ acts on them by $\frac{\pi}{k\tau_2} \frac{\partial }{\partial A_{\bar z}}$. 
If we insert $A_{\bar z}$ into the overlap \eqref{holo overlap}, then on one hand it can act on $\Psi_2(A_{\bar z})$ by just multiplying it by $A_{\bar z}$, but on the other hand by integration by parts it can act on $(\Psi_1(A_{\bar z}))^*$, which is a function of $A_z$, as $\frac{\pi}{k\tau_2} \frac{\partial}{\partial A_z}$, as expected from the canonical commutation relations. 

We will parameterize the value of $A_{\bar z}$ at the boundary by
\bea
A_{\bar z}={i\pi \over \tau_2}\xi, \label{holcor}
\eea
where $\xi$ is a complex number. As described above, we can write the wavefunctions on the torus, in particular $\Psi_{\rm interior}$, as holomorphic functions of $\xi$.
The normalization in \eqref{holcor} has been chosen so that large gauge transformations in the bulk $A\rightarrow A+\omega$, 
which are characterized by integer winding numbers $n,m$ around the two basis cycles of the torus and preserve the gauge of $A$ being constant on the torus,  shift $\xi$ by
\bea
\xi \rightarrow \xi +n+m\tau. \label{LGT}
\eea

Any holomorphic wavefunction of $A_{\bar z}$ gives a ground state of the Hamiltonian, so the only constraint on the wavefunctions comes from their required covariance under large gauge transformations. 
Under these transformations, the interior wavefunction $\Psi_{\rm interior}$ should change as follows
\bea 
\label{pilgt}
\Psi_{\rm interior} \rightarrow \Psi_{\rm interior}\, e^{{ik\over 4\pi}\int\limits_{\partial \mathcal M}\omega \wedge A}
e^{{k\tau_2\over 2\pi}(A_z \omega_{\bar z}+A_{\bar z} \omega_z+|\omega_z|^2)}
e^{i \varphi(\omega)},
 \eea
where we have introduced an additional cocycle $\varphi$ to assure associativity of large gauge transformations
$ \Psi_{\rm interior}(A+(\omega+\omega'))= \Psi_{\rm interior}((A+\omega)+\omega')$.  This condition requires 
\bea
\varphi(\omega+\omega')=\varphi(\omega)+\varphi(\omega')-{k\over 4\pi}\int\limits_{\partial \mathcal M} \omega \wedge \omega',
\eea
understood mod $2\pi$. 
Note that the $A_z$-dependent terms in \eqref{pilgt} cancel, consistent with $\Psi$ being holomorphic in $A_{\bar z}$. 
 
Written explicitly in terms of $\xi$, see Appendix \ref{App:CS} for details, 
\bea
\label{lgt}
\Psi(\xi+n+m\tau)=\Psi(\xi)\, e^{{k\pi\over \tau_2}(n+m\bar\tau)\xi+{k\pi\over 2\tau_2}|n+m\tau|^2+i\varphi},\quad \varphi=\pi k nm+n \phi_1+m\phi_2,
\eea
which is consistent with the combination $|\Psi_{\rm interior}|^2 e^{-{k \pi\over \tau_2}|\xi|^2}$ being invariant under large gauge transformations \eqref{pilgt}, as is expected from \eqref{holo overlap}.
For even $k$ the CS theory does not require a spin structure, and we have $\phi_1=\phi_2=0$. For odd $k$ the definition of the theory requires a spin structure, and on the torus there are four possible spin structures, which give rise to the options $\phi_{1,2} = 0,\pi$. This statement can be justified by considering transformations of $\Psi$ under modular transformations of $\tau$. 

For any choice of spin structure there are $k$ distinct solutions for $\Psi_r(\xi)$ (labeled by $r=0,\cdots,k-1$) since the space of level-$k$ $U(1)$ Chern-Simons wavefunctions on a torus is $k$-dimensional. They can be written explicitly as
\bea
\label{psir}
\Psi_r(\xi)&=&\frac{1}{\eta(\tau)} \sum\limits_n e^{i\pi \tau p^2+2\pi i p u +{\pi u^2\over 2\tau_2}+{\xi(\phi_2-\bar\tau \phi_1)\over 2\tau_2}-{|\phi_2-\tau \phi_1|^2\over 8\pi k \tau_2}}, \\
p&=&\sqrt{k}\left(n+{r\over k}\right),\qquad u=\sqrt{k}\left(\xi+{\tau \phi_1-\phi_2\over 2\pi k}\right). \nonumber
\eea
Here the appearance of $\eta(\tau)$ in the denominator is due to small (topologically trivial) fluctuations of the gauge field in the bulk  \cite{Porrati:2021sdc}. It can be checked straightforwardly that $\Psi_r$ satisfies \eqref{lgt} and is canonically normalized, see Appendix \ref{App:CS}. 

In the holomorphic representation,
the Wilson loop operator $W(p,q)={\rm exp}\,({i\oint_{p,q} A})$ defined along the cycle $p+q\tau$ acts on $\Psi_r(\xi)$ as follows
\bea
\oint_{p,q} {A}&=&(p+q\tau){-i\over k}{\partial \over \partial \xi}+(p+q\bar \tau) {i\pi\over \tau_2}\xi, \label{Wilsonpq}\\
W(p,q)\, \Psi_r(\xi)&=&e^{i(p\phi_1+q\phi_2)/k+2\pi i p r/k+ i\pi pq/k}\Psi_{r+q}(\xi) .
\eea
Note that the spin structure with $\phi_1=\phi_2=\pi$ is one where the spinors are periodic along both basic cycles of the torus; this odd spin structure is modular invariant by itself, but it does not allow any of the cycles to shrink in the interior (so it does not appear for handlebodies). This is consistent with the fact that for $k=1$ with this choice, the unique wavefunction $\Psi_0(\xi)$ has eigenvalues $W(1,0)=W(0,1)=-1$. 

\subsection{Wavefunction of the $(U(1)\times U(1))_k$ theory }
\label{sec:CSU1U1}
Our next step is to study the ``AB'' theory with the action 
\bea
S={ik\over 4\pi}\int (A \wedge dB+B\wedge dA \label{AdB}),
\eea
with invariance under all gauge transformations of $A$ and $B$, that include the
large gauge transformations 
\bea
A\rightarrow A+\omega_A,\quad B\rightarrow B+\omega_B. \label{LGTAB}
\eea
This  defines the theory in the bulk, which we will denote by $(U(1)\times U(1))_k$ to emphasize that this is not a direct product of two $U(1)_k$ Chern-Simons theories. As above, to describe the theory on a handlebody we also need to choose boundary terms and boundary conditions. Unlike the case of a single $U(1)$ field, the $U(1)\times U(1)$ theory has a continuous family of choices of which variables can be kept fixed at the boundary. For any $r$ we can define gauge fields $A_{\pm}= (A/r\pm  B\, r)\sqrt{k/2}$  such that the action becomes 
\bea
\label{ApAm}
S={i\over 4\pi} \int (A_+ \wedge dA_+ - A_- \wedge dA_-).
\eea
This now becomes the action of two decoupled $U(1)$ theories at levels $1$ and $(-1)$, but the dynamical fields are connected at the level of the large gauge transformations \eqref{LGTAB}, so the theory is not equivalent to a product of two decoupled theories. In any case, since $A_+$ has positive level and $A_-$ has negative level, we can now choose boundary terms and boundary conditions as in the previous subsection
such that we fix $(A_+)_{\bar z}$ and $(A_-)_z$ at the boundary to arbitrary values.
Analogously to \eqref{holcor}, we introduce two independent holomorphic coordinates  
\bea
\label{xidef}
\xi={\tau_2\over i\pi}(A_+)_{\bar z},\quad \bar \xi =-{\tau_2\over i\pi}(A_-)_{z},
\eea
and deform the action by the boundary term 
\bea
S'=S-{1\over 2\pi}\int_{\partial M} d^2z \left(|A_+|^2+|A_-|^2\right), \label{bulkABhol}
\eea
such that the equations of motion $\delta S/\delta (A_+)_z=\delta S/\delta (A_-)_{\bar z}=0$ are trivially satisfied at the boundary. The wavefunction $\Psi(\xi,\bar \xi)$ associated with this action is holomorphic in $\xi$, and separately in ${\bar \xi}$.

Next, we demand that under large gauge transformations with parameters $(n,m)$ for $A$, and $(p,q)$ for $B$, which take
\bea \label{large gauge}
\xi &\rightarrow& \xi+\delta \xi, \quad \delta \xi=\sqrt{k\over 2}\left((n+m\tau)r^{-1}+(p+q\tau)r\right),\\ \nonumber
\bar \xi &\rightarrow& \bar  \xi+\delta \bar \xi, \quad \delta \bar \xi= \sqrt{k\over 2}\left((n+m\bar\tau)r^{-1}-(p+q\bar \tau)r\right),
\eea
$\Psi$ should change by
\bea
\Psi&\rightarrow & \Psi\, e^{{\pi\over 2\tau_2}(2 \xi \delta \xi^* + |\delta \xi|^2 + 2\bar \xi \delta \bar \xi^* + |\delta \bar \xi |^2) +i \pi k(mp-nq)}.
\eea 
In this case the cocycle factor is simply $\varphi=\pi k(mp-nq)$, so there is no need to introduce nontrivial phases $\phi_i$. There are $k^2$ wavefunctions given explicitly by 
\bea
\label{psiabu1} 
\Psi_{a,b}(\xi,\bar\xi,\tau)&=&{1\over |\eta(\tau)|^2} \sum_{n,m} e^{i\pi \tau p_L^2-i\pi \bar \tau p_R^2+2\pi i (p_L \xi-p_R \bar \xi)+{\pi\over 2\tau_2}(\xi^2+\bar \xi^2)},\quad 0\leq a,b<k,\\
p_{L,R}&=&\sqrt{k\over 2}\left((n+a/k) r^{-1}\pm(m+b/k)r\right),\quad n,m\in \Z. \nonumber
\eea
We would like to emphasize that \eqref{psiabu1} for different $r$ are different representations of the same $k^2$ bulk wavefunctions, expressed as functions of different variables (corresponding to the specific choice of boundary conditions we made). The result \eqref{psiabu1} resembles the partition function of a free scalar CFT, and we will discuss the precise relation in the next section.

Wilson loops of $A$ along the $n+m\tau$ cycle,  $W_A(n,m)={\rm exp}\, (i\oint_{n,m} A)$, with a similar definition for Wilson loops of $B$, act on \eqref{psiabu1} as follows
\bea
\label{WA}
W_A(n,m) \Psi_{a,b}= \Psi_{a,b+m}\, e^{2\pi i a n/k},\\
W_B(n,m) \Psi_{a,b}= \Psi_{a+m,b}\, e^{2\pi i  b n/k}. \label{WB}
\eea
In particular, Wilson lines of both $A$ and $B$ along the $1$ cycle act on $\Psi_{0,0}$ trivially, so $\Psi_{0,0}$ is a consistent wavefunction on thermal AdS -- the handlebody where this cycle shrinks in the interior.

\subsection{General case}
\label{sec:CSgencase}
Before we proceed with the general $(U(1) \times U(1))^n$ theory we would like to  revisit the $U(1)\times U(1)$ case, but instead of starting with the ``AB'' theory \eqref{AdB}, we can start with \eqref{ApAm} and \eqref{xidef} and just impose  large gauge transformations generalizing \eqref{large gauge}
\bea
\label{lgtL}
\delta\left(\begin{array}{c}
\xi +\bar \xi^* \\
\xi -\bar \xi^* \end{array}\right)=\sqrt{2} \Lambda (\vec{n}+\vec{m}\tau),\quad \vec{n},\vec{m}\in \Z^2.
\eea
Here $\Lambda$ defines an even self-orthogonal lattice in $\R^{1,1}$  as in \eqref{Ldef}.  The holomorphic functions of $\xi$ and $\bar \xi$
\bea
\label{generalpsic}
\Psi_{\vec c}(\xi,\bar \xi ,\tau)=\frac{1}{|\eta(\tau)|^2} \sum_{\vec{\ell}}  e^{i\pi \tau p_L^2-i\pi \bar \tau p_R^2+2\pi i (p_L \xi-p_R \bar \xi)+{\pi\over 2\tau_2}(\xi^2+\bar \xi^2)},\\
\left(
\begin{array}{c}
p_L+p_R\\
p_L-p_R\\
\end{array}\right)=\sqrt{2} \Lambda(\vec{\ell}+g_\Lambda^{-1}\vec{c}),\quad \vec{\ell}\in \Z^2 \nonumber
\eea
are parametrized by elements of the Abelian group $\vec{c}\in \Z^2/g_\Lambda=\Lambda^*/\Lambda={\sf G}$, and under large gauge transformations \eqref{lgtL} 
they change as follows :
\bea
\Psi_{\vec c}(\xi +\delta \xi,\bar \xi+\delta \bar\xi)=\Psi_{\vec c}(\xi, \bar\xi)\, e^{{\pi \over 2\tau_2}(2 \xi \delta \xi^*+ |\delta \xi|^2 +2 \bar \xi \delta \bar \xi^* + |\delta \bar \xi |^2)}\,e^{i\pi n^T g_\Lambda m }.
\eea

The generalization to the case of $(U(1) \times U(1))^n$ is now straightforward. The main ingredient is the even self-orthogonal lattice ${\Lambda} \in \R^{n,n}$, which defines large gauge transformations of  $U(1)^n$-valued gauge fields $A_\pm$ as follows, 
\bea
\label{lgtXI}
\left(\begin{array}{c}
\xi +\bar \xi^* \\
\xi -\bar \xi^* \end{array}\right)\rightarrow 
\left(\begin{array}{c}
\xi +\bar \xi^* \\
\xi -\bar \xi^* \end{array}\right) +{\sqrt{2} {\Lambda} (\vec{n}+\vec{m}\tau)},\quad \vec{n},\vec{m}\in \Z^{2n},
\eea
while the relation between $\xi,\bar \xi$ and $A_\pm$ is as in \eqref{xidef}, generalized to vector-valued quantities. The resulting wavefunction is parametrized by an element of  the Abelian group $c\in {\sf G}_{\Lambda}={\Lambda}^* /{\Lambda}$, 
\bea
\label{Psig}
&&\Psi_{c}={\Theta_c\over |\eta(\tau)|^{2n}}, \nonumber \\
&&\Theta_{c}(\xi,\bar \xi ,\tau)=\sum_{\vec{\ell}} e^{i\pi \tau p_L^2-i\pi \bar \tau p_R^2+2\pi i (p_L \cdot  \xi-p_R\cdot  \bar \xi)+{\pi\over 2\tau_2}(\xi^2+\bar \xi^2)},\\
&&\left(\begin{array}{c}
p_L+p_R\\
p_L-p_R\end{array}\right)={\sqrt{2} {\Lambda}(\vec{\ell}+g_{\Lambda}^{-1}\vec{c})},\quad \vec{\ell}\in \Z^{2n},\quad g_{\Lambda}={\Lambda}^T \eta {\Lambda}. \nonumber
\eea
The wavefunction $\Psi_{c}$  coincides exactly with the ``codeword blocks'' \eqref{Psic}. We will explore the holographic interpretation of this result in section \ref{sec:hol}.

\section{Holographic description of the ensemble of code CFTs}
\label{sec:hol}

\subsection{Level $k=1$ CS theories and  conventional holographic correspondence}
\label{sec:convhol}

As discussed above, for $k=1$ the $U(1)\times U(1)$ CS theory has a unique wavefunction, which can be written  as $\Psi_{00}(\tau,\xi,\bar \xi)$. It is given by the CS path integral on any handlebody $\mathcal M$ with the appropriate boundary conditions on $\partial M=\Sigma$ of genus one.  Our starting point is the observation that this unique wavefunction \eqref{psiabu1} with $k=1$ is the same as the {\it path integral} of the two-dimensional CFT, the compact scalar of radius $R=\sqrt{2}r$, coupled to an external complex gauge field $\mathcal A$ parametrized by  $\xi, \bar \xi$
\bea
Z_{BPI}(\tau,\xi,\bar \xi)=\Psi_{00}(\tau,\xi,\bar \xi).
\eea We discuss the compact scalar in detail in  Appendix \ref{App:Z}. From the bulk point of view $\xi, \bar \xi$ parametrize certain components of the fields $A_+,A_-$ on the boundary of $\mathcal M$, and the holographic dictionary relates them to sources in the CFT by
\begin{eqnarray}
{i \pi\over \tau_2}\xi={\mathcal A}_{\bar z}=(A_+)_{\bar z},\qquad 
-{i \pi\over \tau_2}\bar \xi={\mathcal A}_z=(A_-)_z,
\end{eqnarray}
such that $Z_{BPI}$ is the CFT path integral with these sources (as discussed in Appendix \ref{App:Z}).
In the CFT the complex field $\mathcal A$ is a combination of two real fields $A$ and $B$ coupled to the two conserved $U(1)$ currents, see \eqref{complexA}, and we have chosen notation such that the CFT source fields $A,B$ are exactly the boundary values of the bulk gauge fields $A,B$ introduced in subsection \ref{sec:CSU1U1}. At the same time we emphasize that the Chern-Simons theory is quantized with the boundary condition that fixes the fields $(A_+)_{\bar z}, (A_-)_z$ at the boundary, while
$(A_+)_{ z}, (A_-)_{\bar z}$ vary freely. This condition looks cumbersome when expressed in terms of $A$ and $B$. 

While preserving the same boundary conditions, we can add an additional boundary term ${\pi \xi \bar \xi\over \tau_2}$  to the bulk action \eqref{bulkABhol} and obtain, after an integration by parts,
\bea
S_1={i\over 2\pi}\int_{\mathcal M}B\wedge dA -  {r^2 \over \pi} \int_{\partial \mathcal M}d^2z\, |B|^2.
\eea
The new action still satisfies $\delta S_1/\delta (A_+)_z=\delta S_1/\delta (A_-)_{\bar z}=0$ at the boundary, because the added boundary term does not include fluctuating fields. It leads to a holomorphic bulk wavefunction which equals the first path integral introduced  in Appendix \ref{App:Z}
\bea
\label{H1}
Z_{\rm PI}(\tau,\xi, \bar \xi)=\int {\mathcal D}A\, {\mathcal D}B\,  e^{-S_1}.
\eea
Note that unlike the bulk path integral discussed in the previous section (related to it by $Z_{PI}(\tau,\xi,\bar \xi) = Z_{BPI}(\tau,\xi,\bar \xi) e^{-{\pi \over \tau_2} \xi \bar\xi}$), $Z_{\rm PI}$ is manifestly invariant under large gauge transformations of $A$. Similarly, subtracting ${\pi \xi \bar \xi\over \tau_2}$ from the bulk action leads to 
\bea
S_2={i\over 2\pi}\int_{\mathcal M}A\wedge dB -  {r^{-2}\over \pi} \int_{\partial \mathcal M}d^2z\,  |A|^2,
\eea
and 
\bea
\label{H2}
Z'_{\rm PI}(\tau,\xi, \bar \xi)=\int {\mathcal D}A\, {\mathcal D}B\,  e^{-S_2},
\eea
which is manifestly invariant under large gauge transformations of $B$. To be precise, 
these bulk theories are actually complexifications of the path integrals considered in Appendix \ref{App:Z}, as $\xi, \bar \xi$ are treated here as two independent complex variables. To reduce precisely to the CFTs described by the actions \eqref{S1} and \eqref{S2}, one would need to impose additional restrictions on the boundary conditions in \eqref{H1} and \eqref{H2}, that respectively set $B=0$ and  $A=0$ at the boundary, by choosing $\xi^*=\pm \bar \xi$. 

It is important to point out that the parameter $r$ from the bulk point of view is a parameter changing the representation of the Chern-Simons wavefunction $|\Psi\rangle$. It defines the boundary conditions, but it does not  affect the action in the bulk. Hence, for all $r$ the quantum state $|\Psi\rangle$ remains the same, which is already clear from the fact that the Hilbert space of level-1 Chern-Simons theory is 1-dimensional. From the boundary point of view, the parameter $r$ -- the radius of the compact circle -- changes the CFT. This situation is not conceptually different from more conventional instances of the holographic correspondence, such as gauge/gravity duality, in which the path integral of the same bulk action with different boundary conditions describes different field theories. For example, exactly marginal deformations of a CFT correspond to changing the boundary conditions for massless scalars in anti-de Sitter space. Another even more direct analogy is with the dS/CFT correspondence \cite{Strominger:2001pn}, where the same quantum Hartle-Hawking wavefunction in the bulk is dual to different field theories, depending on the overlap of the unique bulk wavefunction with the wavefunction at the boundary. 

The generalization of the above considerations to  $(U(1) \times U(1))^n$ is straightforward.  The general level $k=1$ theory is $(U(1) \times U(1))^n$ Chern-Simons theory,  quantized with large gauge transformations corresponding to points in an even self-dual lattice $\Lambda$  \eqref{lgtXI}.  In this case ${\sf G}_{\Lambda}={\Lambda}^*/{\Lambda}$ consists of a single element, and the unique wavefunction 
\eqref{Psig} is identified with the path integral of the Narain theory associated with $\Lambda$
\bea
\label{holography}
Z_{BPI}^{}(\tau,\xi,\bar \xi, {\Lambda})=\Psi_{\vec{0}}(\tau,\xi,\bar \xi).
\eea 
Similarly to the $n=1$ case, there are many possible definitions of path integrals, generalizing \eqref{H1} and \eqref{H2}; here we use the $T$-duality invariant definition
of \eqref{Psig}.

One can rewrite the $(U(1) \times U(1))^n$ fields $A_\pm$ in terms of the gauge fields $A,\, B$
and the lattice $\Lambda$, which we parametrize by $\gamma$ and $B$ as in \eqref{LCC} with trivial $O_T$,
\bea \label{gammaB}
A_{\pm}={(\gamma\pm \gamma^* B)A\pm \gamma^* B\over \sqrt{2}}.
\eea
In terms  of these fields, the action of large gauge transformations are canonical, $A\rightarrow A+\omega_A$, $B\rightarrow B+\omega_B$, but the boundary conditions in the path integral become $\Lambda$-dependent. This description provides  the holographic dictionary for a general Narain CFT: the path integral of a  Narain theory is equivalent to the path integral of level-1 $(U(1)\times U(1))^n$ Chern-Simons theory with boundary conditions (wavefunction representation) specified by $\Lambda$. 

The construction above is an explicit realization of the $\rm AdS_3/CFT_2$ correspondence for Narain theories in terms of pure Chern-Simons theory in the bulk. 
The  original treatment in \cite{Gukov:2004id}, recently revisited in \cite{Ashwinkumar:2021kav}, was  in terms of Maxwell-Chern-Simons theory in the limit of infinite Maxwell coupling.  
We have shown here  that inclusion of  the Maxwell term is not necessary, provided a more  explicit treatment by evaluating the path integral on both sides of the duality, and established a holographic dictionary. Our approach  is also related to the recent  work \cite{Benini:2022hzx}, which constructs a bulk description for a Narain theory with decomposable ${\Lambda}={\Lambda}_L \oplus {\Lambda}_R$ in terms of level $k=1$ CS theory, obtained from $k>1$ Chern-Simons theory by gauging all discrete symmetries. 

It is important to note that the holographic description above does \underline{not} include a sum over bulk geometries or topologies.  Rather, all handlebodies yield the same wavefunction, obeying \eqref{holography}. This is analogous to the case of $\rm AdS_3$ with $k=1$ units of NS-NS flux (the ``tensionless string'') \cite{Eberhardt:2020bgq,Eberhardt:2021jvj}, where  fluctuations in the bulk on a fixed background geometry are believed to account for the full partition function. 

Codes, and code ensembles, play a rather trivial role in the holographic description of Narain CFTs in terms of level $k=1$ Chern-Simons theory. Indeed, the $k=1$ $U(1)\times U(1)$ theory  is associated  with the unique length $n=1$ code over the alphabet $\Z_1 \times \Z_1$, consisting of the unique codeword $c=(0,0)$. The parameter $r$, the radius of the compact scalar on the CFT side, does not affect the code, but controls the embedding of the code into the space of $n=1$ Narain CFTs. Similarly,  whenever the  lattice $\Lambda$ is self-dual, the group  ${\sf G}_{\Lambda}$ is trivial, consisting of a single element, and there is a unique code consisting of a single codeword $c=\vec{0}$. As we saw above, this trivial code can be mapped to an arbitrary Narain theory by choosing an appropriate embedding.  To summarize, we see that the conventional holographic correspondence emerges when the code ensemble consist of only one element, a unique code associated with a given Narain CFT.

\subsection{Averaging over Narain CFTs}
\label{sec:bcaveraging}

The holographic duality described above maps Narain theories to three dimensional bulk theories which are non-gravitational Chern-Simons theories living on a fixed space-time, with no sum over geometries. This is consistent with the fact that in these CFTs the energy-momentum tensor is a product of $U(1)$ currents, so we do not expect an independent graviton field in the bulk 

Motivated by \cite{Afkhami-Jeddi:2020ezh,Maloney:2020nni} will now consider an average over a finite ensemble of Narain theories, or over the whole moduli space of Narain theories with some $c=n$. {\it A priori},  it is not clear if such an ensemble average would  have a simple description in the bulk.  The duality between Narain theories and level-1 CS theories on a fixed handlebody provides one way to evaluate it -- by averaging over all possible boundary terms and boundary conditions, corresponding to all Narain CFTs. For $n=1$ this is just an average over the values of $r$. One way to implement this is to write down the boundary terms as a function of $\gamma$ and $B$ using \eqref{gammaB}, and then to make these variables dynamical and to integrate over them with an $O(n,n,\R)$-invariant measure. These variables live just on the boundary, but since they are constant on $\partial {\mathcal M}$, integrating over them gives a non-local theory. This non-local theory (on a given handlebody) is, by construction, equivalent to the ensemble average over Narain theories, but this is not very satisfactory, and certainly more complicated than the dual description suggested in \cite{Afkhami-Jeddi:2020ezh,Maloney:2020nni}. In the rest of this section we explore an alternative way to obtain the ensemble average over Narain theories, which will lead to bulk sums over geometries  similar to  those of \cite{Afkhami-Jeddi:2020ezh,Maloney:2020nni}, but described by a fully consistent Chern-Simons theory with compact gauge group.
``U(1)-gravity'' theory will then emerge as a limit.

\subsection{Level $k>1$ CS theory and ensemble averaging}
\label{sec:Pseries}
Our next step is to consider codes over alphabets with more than one element. As we have seen in full generality in  section \ref{sec:CSgencase},  the ``codeword blocks'' $\Psi_c$ \eqref{Psic} appearing in the context of codes have a simple bulk interpretation. They are precisely the same as the wavefunctions of the $(U(1) \times U(1))^n$ Chern-Simons theory on a spatial torus. Indeed, the theory quantized with large gauge transformations specified by a lattice $\Lambda$ has exactly $G_{\Lambda}={\Lambda}^*/{\Lambda}$ independent wavefunctions, in one to one correspondence with the codewords. As was emphasized in section \ref{sec:codes}, any given code $\CC$ of length $n$ can be  associated with any Narain CFT of central charge $n$, by choosing an appropriate embedding. As a result we have the following expression for the CFT path integral 
\bea
\label{ZPsi}
Z_\CC=W_\CC(\Psi)=\sum_{c\in \CC}\Psi_c.
\eea 
This expression, though suggestive, has no apparent holographic interpretation. Indeed, the sum of wavefunctions on the right-hand side of \eqref{ZPsi} does not allow for a simple interpretation as the 
bulk path integral evaluated on a simple 3d geometry, or as a sum of such path integrals. This is because in general, path integrals on  simple geometries such as solid tori with different shrinking cycles would lead to a subclass of $\Psi_c$ not nearly exhausting all possibilities. This is easiest to see  in the case of  codes over $\Z_k\times \Z_k$ and $(U(1) \times U(1))_k^n$ Chern-Simons theory.  The path integral over the solid torus with a  shrinkable $n+m\tau$ cycle will lead to the unique combination of $\Psi_{a_1, b_1} \dots \Psi_{a_n b_n}$ invariant under (\ref{WA}),(\ref{WB}) for those values of $n$ and $m$.  
For $k>1$, combinations of these with integer coefficients can not in general 
lead to  \eqref{ZPsi}. 

Although associating individual CFTs with codes does not lead to a simple holographic interpretation, we note that codes -- and  hence also the associated CFTs -- naturally appear in the context of ensembles. There is always an ensemble of all codes (CFTs) of a particular type (e.g. over a particular alphabet) and of given length (corresponding to CFT central charge $n$). It was initially suggested in \cite{Dymarsky:2020qom} that such an ensemble of code CFTs should admit a holographic interpretation. 
A crucial observation building towards such an interpretation was made recently in \cite{Henriksson:2022dml}. There the authors considered an ensemble consisting of all length-$n$  $\Z_2\times \Z_2$ codes  with the glue matrix 
\bea
{\sf\Lambda}=\left(\begin{array}{cc}
1 & 1\\
1 & -1\end{array}\right)
\eea
in the notation of section \ref{sec:codes} (the original paper used a different but equivalent description). It was conjectured, and then verified explicitly for small $n$, that the enumerator polynomial, averaged over the ensemble of all such codes, is proportional to the ``Poincar\'e series'' of all possible modular transformations acting on the auxiliary variable associated with the trivial codeword,
\bea
\label{PSeries}
\overline{W}(\{X\})\equiv {1\over {\mathcal N}} \sum\limits_{\CC} W_\CC(\{X\})\, \propto \sum_{ \g \in \Gamma^* \backslash SL(2,\Z)} {\g}(X_{00}^n).
\eea
We emphasize that the action of the modular group on the variables $X_{ab}$, generated by (\ref{ST}), along with the equality \eqref{PSeries}, are defined and satisfied at the level of codes, before the map to CFTs and Chern-Simons theory. The variables $X_{ab}$ provide (for $k=2$) a  four-dimensional representation of the  modular group $SL(2,\Z)$; hence the sum on the right-hand side of \eqref{PSeries} over $\Gamma^*\backslash SL(2,\Z)$, where $\Gamma^*$ is the stabilizer of $X_{00}$, includes a finite number of terms. Alternatively, one can sum over the whole modular group, with the infinite size of $\Gamma^*$ absorbed into the overall proportionality coefficient. 

While we leave a systematic justification of \eqref{PSeries} and its generalizations to future work \cite{future}, we point out that the equality between the  weighted average over codes and the Poincar\'e sum over the modular group will apply to other code constructions as well, and will extend to higher genus boundaries. To be explicit, in what follows we focus our attention on $\Z_k \times \Z_k$ codes with the glue matrix \eqref{Lambdak} and prime $k=p$, for which we establish the analogue of \eqref{PSeries}  for arbitrary $n$. We consider codes of length $n$, which define an ensemble of size \eqref{numberofcodes}. For prime $p$ all codes should be averaged with equal weight.  In this case, the average is straightforward -- see footnote on page 31. 

Next we consider the sum over the modular group in \eqref{PSeries}. 
We  introduce $p^2$ variables $X_{ab}$ forming a representation of the modular group
generated by \eqref{Ev} and by obtaining $X'$ of \eqref{MW} when taking $\tau \to -1/\tau$. The full set of $(p^2)^n$ variables $X_{\vec{a},\vec{b}}$ 
associated with codes of length $n$ transform in the tensor product of $n$ such representations. The explicit action of the $T$ and $S$ generators of $SL(2,\Z)$ is, using (\ref{ST}), 
\bea
&& T(X_{\vec{a},\vec{b}})=X_{\vec{a},\vec{b}}\, e^{2\pi i {\vec{a}\cdot \vec{b}\over p}}
,\\
&& S(X_{\vec{a},\vec{b}})={1\over p^n}\sum_{\vec a',\vec b'} X_{\vec{a}',\vec{b}'}\, e^{-2\pi i {\vec{a}\cdot \vec{b}'+\vec{a}'\cdot \vec{b}\over p}}. \nonumber
\eea
Our goal is to sum over $\Gamma^*\backslash SL(2,\Z)$, where $\Gamma^*$ is the stabilizer group of $X_{\vec{0},\vec{0}}$.  The sum can be performed in two steps: we first define a subgroup  $\Gamma\subset SL(2,\Z)$ which leaves all $X_{\vec{a},\vec{b}}$ invariant, and then additionally factorize over the stabilizer of $X_{\vec{0},\vec{0}}$ within $\Gamma\backslash SL(2,\Z)$. The group $\Gamma$ in the general case is known to be a congruence subgroup of $SL(2,\Z)$ \cite{Ashwinkumar:2021kav,schoeneberg2012elliptic}; for prime $p$ it is 
the principal congruence subgroup $\Gamma=\Gamma(p)$. The stabilizer of $X_{\vec{0},\vec{0}}$ in $\Gamma\backslash SL(2,\Z)$ is the cyclic group $\Z_2\times \Z_p$ generated by $S^2$ (which takes $X_{a,b}$ to $X_{-a,-b}$) and by powers of  $T$, so 
\bea
\Gamma^* \backslash SL(2,\Z)=(\Z_2\times \Z_p) \backslash SL(2,\Z)/\Gamma(p).
\eea
This quotient consists of $(p^2-1)/2$ elements, which can be parametrized by integer pairs $(c,d)\sim (-c,-d)$, corresponding to the modular transformation
\bea
g=\left(\begin{array}{cc}
a & b\\
c & d\end{array}\right)\in SL(2,\Z_p)\cong\Gamma_1(p)\backslash PSL(2,\Z),
\eea
which has the following action on $X_{\vec{0},\vec{0}}$:
\bea
\label{modularPsi}
g(X_{\vec{0},\vec{0}})={1\over p^n}\sum_{\vec{a},\vec{b}} X_{\vec{a},\vec{b}} \, e^{-2\pi i {\vec{a}\cdot \vec{b}\over p}r},\qquad r=d/c\, \,{\rm mod}\, \, p.
\eea
The equation above applies when $c\neq 0$; otherwise $g(X_{\vec{0},\vec{0}})=X_{\vec{0},\vec{0}}$.
One can readily see that the $(p^2-1)/2$ terms in the Poincar\'e sum split into $p+1$ terms labeled by elements of $\Gamma_0(p)\backslash SL(2,\Z)=\{1,ST^l\}$ with $0\leq l<p$, each appearing $(p-1)/2$ times. Combining everything, we find the averaged enumerator polynomial for codes over $\Z_p \times \Z_p$, 
\bea
\nonumber
\overline{W}(X_{\vec{a},\vec{b}})={1\over {\mathcal N}} \sum\limits_{\CC} W_\CC(X_{\vec{a},\vec{b}})={\sum_{g \in \Gamma_0(p)\backslash SL(2,\Z)}g(X_{\vec{0},\vec{0}})\over 1+p^{1-n}}={X_{\vec{0},\vec{0}}+{1\over p^{n}}\sum\limits_{r=0}^{p-1} \sum\limits_{\vec{a},\vec{b}}X_{\vec{a},\vec{b}}\, e^{-2\pi i {\vec{a}\cdot \vec{b}\over p}r}\over 1+p^{1-n}}.\\
\label{main}
\eea
Here $\mathcal N$ is given by \eqref{numberofcodes}, and
the coefficient $1+p^{1-n}$ in the denominator of the right-hand side  is chosen such that the coefficient in front of $X_{\vec{0},\vec{0}}$ associated with the trivial codeword is equal to one. 

The identity \eqref{main} for code CFTs acquires a straightforward holographic interpretation which has  been envisioned in \cite{Henriksson:2022dml}. We consider $((U(1) \times U(1))_k^n$ Chern-Simons theory, placed on an arbitrary handlebody geometry,
and  identify $\Psi_{\vec{0},\vec{0}}(\tau, \xi, \bar \xi)$ to be the wavefunction of this theory on thermal AdS, the solid torus with shrinkable $a$-cycle. Indeed, Wilson loops of both $A$ and $B$ fields over this shrinkable cycle should act on the boundary wavefunction trivially, which singles out $\Psi_{\vec{0},\vec{0}}$, as follows from (\ref{WA}),\,(\ref{WB}). 
Hence, the sum in \eqref{main} can be interpreted as a sum over all possible handlebody topologies, or more accurately, as a sum over equivalence classes of topologies yielding the same boundary wavefunction,
\bea
\label{ensembleH}
\overline{Z}_{BPI}(\tau,\xi, \bar \xi)&=&{1\over 1+p^{1-n}}\sum_{\g\in \Gamma_1(p) \backslash SL(2,\Z)} \Psi_{\vec{0},\vec{0}}(\g\, \tau, \g\, \xi, \g\, \bar \xi)=
\\
&&{1\over 1+p^{1-n}} \left(\Psi_{\vec{0},\vec{0}}(\tau,\xi, \bar \xi) +p^{-n}\sum\limits_{r=0}^{p-1} \sum\limits_{\vec{a},\vec{b}\in \Z^n_p} \Psi_{\vec{a},\vec{b}}(\tau,\xi, \bar \xi)\,  e^{-2\pi i {\vec{a}\cdot \vec{b}\over p}r}\right). \nonumber
\eea
This equality between ensemble averaging 
over code CFTs on the field theory side, 
and summing over topologies on the bulk side, is preserved under the action of $O(n,n,\R)$, which changes the embedding of the codes in the space of Narain CFTs and the representation of the wavefunction   of the dual Chern-Simons theory.

We expect an  equality analogous to \eqref{ensembleH} to apply more broadly, to codes  defined as even self-dual linear subspaces of the group ${\sf G}_{\Lambda}$ defined by an even self-orthogonal lattice ${\Lambda}$ and  corresponding code CFTs/ Chern-Simons theories, although the averaging weights and the details of the ``Poincar\'e sum'' will be different. The equality will also extend to higher genus boundary geometries. 

To summarize, 
we have obtained an infinite series of explicit examples of ``holographic duality,''  in which an ensemble of CFTs is dual to a Chern-Simons theory coupled to topological gravity, and the bulk partition function is given by a sum over topologies. This  sum  is akin to a sum over saddle point configurations in conventional gravity, implementing the ideas of \cite{Dijkgraaf:2000fq,Maloney:2007ud,Keller:2014xba}. In spirit, our examples are similar to
the original work \cite{Castro:2011zq} representing the Ising model partition function as a sum ``over geometries,'' but crucially we  explicitly outline the dual theory in the bulk. 

Our code-based construction allows for many generalizations, to additive codes of other types, and potentially going beyond additive codes and Abelian CFTs. We expect that this approach may lead to many more explicit examples, potentially reformulating the results of \cite{Meruliya:2021lul,Dong:2021wot,Raeymaekers:2021ypf} in terms of codes. 

Although in this paper we only consider the CFTs living on a torus, we expect that the holographic duality will generalize to higher genera \cite{future}.
The ensembles we consider contain a finite number of CFTs, hence the ensemble is parameterized by a finite number of moments.  This will imply that the dual Chern-Simons theory on $\mathcal M$,  with $\partial \mathcal M$ being a Riemann surface of sufficiently high genus, would be completely determined in terms of path integrals over lower genus boundary surfaces. This will require various factorization rules in the bulk, which deserves a better understanding. 

\subsection{Holographic correspondence in the $k \rightarrow \infty$ limit}
\label{sec:kinf}

In the previous section we saw that the size of the ensemble is related to the number of classes of topologies appearing in the bulk sum. Bigger ensembles correspond to ``more gravitational'' theories in the bulk, distinguishing more topological features and thus leading to a sum over more classes of topologies. It is thus natural to ask, what would happen with the duality as the size of the  code CFT ensemble becomes infinitely large.
In what follows we take $k=p$ to be prime for simplicity.   To evaluate the right-hand side of \eqref{main} in the $p\to \infty$ limit we go back to definition of $\Psi_{\vec{0},\vec{0}}$ \eqref{Psig},
\bea
\Psi_{\vec{0},\vec{0}}(\tau',  \xi', \bar \xi')={\Theta_{\vec{0},\vec{0}}(\tau',\xi', \bar \xi')\over |c\tau+d|^n |\eta(\tau)|^{2n} },\qquad 
\g=\left(\begin{array}{cc}
a & b\\
c & d\end{array}\right)\in SL(2,\Z),\\
\tau'={a\tau+b\over c\tau+d},\qquad  \xi'={\xi\over c\tau+d},\qquad \bar \xi'={\bar \xi\over c\bar\tau+d}. \nonumber
\eea
From \eqref{modularPsi} we know that we do not need to consider all possible co-prime pairs $c,d$, but only $c=0,d=1$, and $p$ additional pairs yielding all possible values for $d c^{-1}$ mod $p$. A crucial observation is that one can always pick a set of such pairs with positive $c$ satisfying $c,|d|\leq \sqrt{p}$. While we could not prove this in full generality, we have numerically checked it for the first hundred primes. 

We first consider the case when $c,|d|\ll \sqrt{p}$ with $p\gg 1$. In this regime all vectors $\vec{p}_L,\vec{p}_R$ summed over in $\Theta_{\vec{0},\vec{0}}$, except for $\vec{p}_L,\vec{p}_R=0$, are of order 
  $|\vec{p}_L|, |\vec{p}_R| \sim O(p^{1/2})$. This is in fact a general result for any embedding, in the limit where the embedding is fixed while $p\rightarrow \infty$. So for $c,|d|\ll \sqrt{p}$, the factor 
\bea
e^{-\pi \tau'_2 (|\vec{p}_L|^2+|\vec{p}_R|^2)}
\eea
in \eqref{Psig} suppresses all other terms, and hence the only contribution is from $\vec{p}_L,\vec{p}_R=0$,
\bea
\Theta_{\vec{0},\vec{0}}(\tau',\xi', \bar \xi')=e^{{\pi\over 2\tau'_2}(\xi'^2+\bar \xi'^2)}+e^{-O(p)}.
\eea
Outside of the $c,|d|\ll \sqrt{p}$ regime, but provided  $c,|d|\leq \sqrt{p}$ is still satisfied, we notice that the combination $\tau_2' (|\vec{p}_L|^2+|\vec{p}_R|^2)$ is at least of order one, and
$|p_L\cdot \xi|$, $|p_R\cdot \bar \xi|$ are at most of order one, $O(p^0)$. This means 
that $\Theta_{\vec{0},\vec{0}}(\tau',\xi', \bar \xi') \lesssim O(p^0)$ for $p\gg 1$. 
Now, going back to the sum over $p+1$ pairs $c,d$, we split the sum into two groups, for $c,|d|$ satisfying $c,|d|\leq p^\alpha$ for any $1/3<\alpha<1/2$, and the rest. 
The first group will yield 
\bea
\overline{Z}_{BPI}\approx
{1\over 1+p^{1-n}}\sum_{\substack{(c,d)=1, \\
c,|d|\leq p^\alpha}} {e^{{\pi \xi^2 \over 2\tau_2}{c\bar \tau+d\over c \tau+d} +
{\pi \bar\xi^2 \over2 \tau_2}{c \tau+d\over c\bar \tau+d}}\over |c\tau+d|^n |\eta(\tau)|^{2n}},\qquad p\gg 1,
\eea
while the second group, which has at most $p$ terms, will give a contribution bounded by 
\bea
\sum_{|c|+|d|\geq p^\alpha} {\Theta_{\vec{0},\vec{0}} \over |c\tau+d|^n |\eta(\tau)|^{2n}}\lesssim O(p^{1-n\alpha}).
\eea
For $n\,\alpha>1$ this second term is negligible in the limit $p\rightarrow \infty$.  To conclude, for $n\geq 3$, in the $p\rightarrow \infty$ limit we recover the following expression for the averaged partition function 
\bea
\label{flavor}
\overline{Z}(\tau, \xi, \bar \xi)={1\over |\eta(\tau)|^{2n}}\sum_{(c,d)=1} {e^{-i\pi { c\xi^2\over c\tau+d}+i\pi  {c\bar\xi^2\over c\bar \tau+d} }\over |c\tau+d|^n},
\eea
matching the result of \cite{Datta:2021ftn}, which reduces to the partition function of \cite{Maloney:2020nni,Afkhami-Jeddi:2020ezh} for $\xi=\bar\xi=0$. The special cases of $n=1,2$ are considered below. 

Our final expression \eqref{flavor} is manifestly independent of the embedding. From   \cite{Afkhami-Jeddi:2020ezh,Maloney:2020nni,Datta:2021ftn} we know this expression is equal to the Narain CFT path integral averaged with the Haar measure over the whole Narain moduli space. It is thus natural to speculate that for $n>2$ in the $p\rightarrow \infty$ limit, independently of the embedding, the ensemble of code CFTs densely covers the whole moduli space with the canonical measure. We will first discuss how this works in the $n=2$ case in next section, and then provide additional arguments and formulate an underlying hypothesis in section \ref{sec:typicality}.

The original works \cite{Afkhami-Jeddi:2020ezh,Maloney:2020nni} identified the sum in the right-hand-side of \eqref{flavor} as a sum over handlebody topologies of the ``perturbative sector of Chern-Simons,'' an Abelian Chern-Simons theory with only small (topologically trivial) fluctuations of gauge fields contributing to the path integral. This theory was dubbed ``U(1)-gravity''  \cite{Afkhami-Jeddi:2020ezh}, but apparently it has no well-defined microscopic description. As was pointed out in \cite{Maloney:2020nni},  genuine Chern-Simons theories with either non-compact  or compact gauge groups would lead to different results. We are now ready to clarify this point.  U(1)-gravity does not have a proper microscopic description in the bulk because it emerges as a limit of well-defined theories, namely the $k\rightarrow \infty$ limit of level-$k$ 
$(U(1)\times U(1))^n$ Chern-Simons theory, coupled to topological gravity (to give the sum over handlebodies).  

\subsection{Ensembles of $n=1$ and $n=2$ theories in the large $p$ limit}
 \label{c=1c=2}

The cases $n=1$ and $n=2$ are special. As discussed above, for $n=1$ and $k>1$ the ensemble consists of just two codes, one with the codewords of the form $(a,0)\in \CC_1$ and the other with $(0,b)\in \CC_2$, with arbitrary $0\leq a,b<p$. When translated to CFTs, each of them maps to a single compact scalar, with radii $R_{\pm}=\sqrt{2}\, r\, p^{\pm1/2}$, respectively. The relation between the ensemble-averaged enumerator polynomial and the Poincar\'e series \eqref{main} is valid for all $n$; hence this ensemble can be represented ``holographically" as follows
\bea
\label{n=1}
Z_{BPI}(\tau, \xi, \bar \xi, R_+)+Z_{BPI}(\tau, \xi, \bar \xi, R_-)=\sum_{\g \in \Gamma_0(p) \backslash SL(2,\Z)} \Psi_{00}(\g\, \tau, \g\, \xi, \g\, \bar \xi, r),
\eea
where the sum is over $p+1$ classes of three-dimensional topologies and $\Psi_{00}$ is given by \eqref{psiabu1}.
We have explicitly specified the embedding parameter $r$, which is arbitrary and can scale with $p$. This relation holds for any prime $p$, but in the limit $p\rightarrow \infty$ it diverges.  The sum in the right-hand side of \eqref{n=1} becomes the divergent real Eisenstein series of weight $1$.
The left-hand side of \eqref{n=1} also diverges, as at least one of the scalars decompactifies.  Using the representation \eqref{rep} of the partition function and T-duality, we find the  limit for fixed $r$ to be (for simplicity we consider vanishing  fugacities $\xi=\bar\xi=0$)
\bea
Z_{R_+}+Z_{R_-}={p\left(r+r^{-1}\right)\over \sqrt{\tau_2} |\eta(\tau)|^2}+e^{-O(p)},\qquad p\rightarrow \infty.
\eea

One can also consider the $p\to \infty$ limit when $R_-=R$ remains fixed, while $R_+$ scales with $p$, in which case 
\bea
Z_{R}+Z_{R p}= {(p+1) R\over \sqrt{2\tau_2} |\eta(\tau)|^2}+\sum_{(c,d)=1} R {
\theta_3 \left({i R^2\over 2\tau_2'}\right)-1
\over \sqrt{2\tau_2} |\eta(\tau)|^2}+e^{-O(p)},\quad  \tau_2'={\tau_2\over |c\tau+d|^2}.
\eea
This is essentially the ``modular sum'' representation for the given compact scalar partition function $Z_R$, except for the constant term $R/\sqrt{2\tau_2} |\eta(\tau)|^2$. In this way it is similar to the representation (3.14) of \cite{Benjamin:2021ygh}. In any case, the interpretation of the divergent equations that arise in these cases is not clear.

The case of $n=2$ codes is similar but much richer. Narain lattices in $\R^{2,2}$ are conventionally parametrized by two modular parameters $t=t_1+it_2$ and $b=b_1+ib_2$, related to $\gamma$ and the $B$-field by
\bea
\label{tbdef}
\gamma=\sqrt{{b_2\over t_2}}\left(\begin{array}{cc}
1 &t_1 \\ 0 & t_2\end{array}\right),\qquad B={b_1}\left(\begin{array}{cc}
0 & 1 \\ -1 & 0\end{array}\right). \label{gamma2}
\eea
T-duality acts on $(t,b)$ as $SL(2,\Z)\times SL(2,\Z)$ with an additional $\Z_2$ exchanging $t \leftrightarrow b$. We denote the partition function of a $c=2$ Narain theory by $Z_{c=2}(\tau,t,b)$. It is modular invariant with respect to all three variables. One can also introduce the partition function of primaries, $\hat{Z}(\tau,t,b)=\tau_2 |\eta(\tau)|^4 Z_{c=2}(\tau,t,b)$, where here and in what follows we assume the fugacities vanish $\xi=\bar\xi=0$. The partition function of primaries remains modular invariant under $\tau$, and exhibits triality -- full symmetry under permutation of its three arguments \cite{dijkgraaf1988moduli, Benjamin:2021ygh}.

There are $2(p+1)$ $n=2$ codes, see Appendix \ref{sec:n=2codes} for a detailed discussion. If we choose an embedding, an orthogonal matrix from $O(2,2,\R)$ introduced in the end of section \ref{sec:codes}, in the form 
\bea
\left(\begin{array}{cc}
\gamma^* & \gamma^* B\\
0 & \gamma\end{array}\right)\in O(2,2,\R),
\eea
parametrized by two modular parameters $t=-1/t_0, b=b_0$, then 
the $2(p+1)$ self-dual lattices of the code theories (after appropriate T-dualities) are explicitly given by
\bea
\label{f1}
\{ t={k+t_0\over p}\quad {\rm or}\quad t=p\, t_0 \} \quad {\rm with}\quad b=b_0,
\eea
and 
\bea
\{b={k+b_0\over p}\quad {\rm or}\quad b=p\, b_0 \} \quad {\rm with}\quad t=t_0,
\eea
where $0\leq k<p$. It is convenient to represent the average over code theories in terms of Hecke operators $T_p$, defined as follows \cite{koblitz2012introduction}. For a modular form $f(\tau)$ of weight $k$ and prime $p$ 
\bea
\label{Hecke}
T_p f(\tau) \equiv p^{k-1} f(p\tau) + {1\over p} \sum_{r=0}^{p-1} f\left(\frac{\tau+r}{p}\right).
\eea
Then, the average over codes is simply
\bea
{p\over 2(p+1)}\left( T_p^t\, Z_{c=2}(\tau,t,b)+T_p^b\, Z_{c=2}(\tau,t,b)\right),
\eea
where we introduced an upper index to indicate the variable that each Hecke operator is acting on. 

The ``sum over topologies''  in the right-hand side of \eqref{main}
\bea
\label{bulkn}
{\Psi_{\vec{0},\vec{0}}(\tau)+p^{-n} \sum_{r=0}^{p-1} \sum_{a,b} \Psi_{\vec{a},\vec{b}}(\tau) \, e^{2\pi i r {\vec{a}\cdot \vec{b}\over p}} \over 1+p^{1-n}}
\eea
can also be simplified for general $n$. Starting from an $O(n,n,\R)$ matrix ${\mathcal O}$ specifying a Narain lattice, the partition function of the corresponding Narain theory can be written as (\ref{Psic})
\bea
&&Z_{\mathcal O}={\Theta(\tau)\over |\eta(\tau)|^{2n}},\quad \Theta(\tau)=\sum_{\vec{\ell}} e^{i\pi \tau p_L^2-i\pi \bar \tau p_R^2},\\
&&\left(
\begin{array}{c}
p_L+p_R\\
p_L-p_R\\
\end{array}\right)= {\mathcal O} \sqrt{2}\, \vec{\ell},\quad \vec{\ell}\in \Z^{2n}. \nonumber
\eea
Now, going back to \eqref{bulkn}, we notice that it can be rewritten as 
\bea
{\Theta(p\tau)+p^{-n} \sum_{r=0}^{p-1} \Theta((\tau+r)/p)\over (1+p^{1-n})|\eta(\tau)|^{2n}}={T_p\, \Theta\over (p^{n-1}+1)|\eta(\tau)|^{2n}}, \label{bulknn}
\eea
where $\Theta$ is defined with the same $\mathcal O$ as the embedding matrix of codes, introduced at the end of section \ref{sec:codes}. The last step in \eqref{bulknn} is justified because $\Theta(\tau)$ is a modular form of weight $n$. Using the definition \eqref{Hecke} we can express $T_p$ acting on a modular form  $f$ of weight $n$ in terms of its action on the modular invariant $\tau_2^{n/2} f$,
\bea
{\tau_2^{n/2}\, T_p\, f }=p^{n/2}\, T_p (\tau_2^{n/2} f).
\eea
Going back to the $n=2$ case and noting that $\tau_2\, \Theta=\hat{Z}(\tau,t,b)$ is exactly the partition function of primaries, we can now rewrite the identity \eqref{ensembleH} as follows:
\bea
\label{bulkboundaryn2}
{p\, (T_p^t\, \hat{Z}+T_p^b\, \hat{Z})\over 2(p+1)\tau_2 |\eta(\tau)|^4}={p\, T_p^\tau\, \hat{Z}\over (p+1)\tau_2 |\eta(\tau)|^4}.
\eea 
In fact a stronger identity holds for any prime $p$, see Appendix \ref{app:n=2},
\bea
\label{triality}
T_p^\tau\, \hat{Z}=T_p^t\, \hat{Z}=T_p^b\, \hat{Z},
\eea
which extends the triality -- permutation symmetry of $\hat Z$ with respect to its arguments. 

In the limit $p\rightarrow \infty$, the points $t={k+t_0\over p}$ form a dense line close to $t_2=0$, that crosses infinitely many copies of the fundamental domain. Once these $p$ points are mapped back to the standard keyhole  domain of $SL(2,\Z)$, they will cover it densely with the standard covariant measure $d^2t /t_2^2$. The contribution of the point $t=p\, t_0$ in the full average will be $1/p$ suppressed, and can be neglected.  Thus, the average over code theories when $p\rightarrow \infty$, at least at leading order, would plausibly approach the average over the fundamental domain of $t$ with the $SL(2,\Z)$ covariant measure, plus the same average over $b$ (note that this is not the same as averaging over all Narain theories).  
Similarly,  thanks to \eqref{bulknn}, the ``bulk'' sum in the $p\rightarrow \infty$ limit will be proportional to the average of $\hat{Z}$ over the fundamental domain of $\tau$ with the  measure $d^2\tau/\tau_2^2$. The same conclusion is supported by the  general result  of \cite{clozel2001hecke,GoldsteinMayer}, that  in the limit $p\to \infty$, for any square-integrable modular function $f$, $T_p(f)$ approaches the integral of $f$ over the fundamental domain $\mathcal F$ with the canonical measure
\bea
\label{clozel}
\left|\left| T_p(f) - \int_{\cal F} f\, d\mu \right|\right|  <  ||f||\,  O(p^{-9/28+\epsilon}) \to 0,
\eea
for large $p$,  where $\epsilon$ is any real number $>0$, and $||f||$ is the Weil-Petersson norm of $f$ \cite{GoldsteinMayer}. The caveat  here is that in our case $\hat{Z}$ is not square-integrable, and the integral over the fundamental domain diverges. We therefore {\it conjecture} that in the $p\rightarrow \infty$ limit, $T^x_p(\hat{Z})$ for $x=\tau$, $t$ or $b$ would be given by the regularized average over the fundamental domain of $x$, {\it plus} an $x$-dependent term, which will {\it not} be dependent on other variables.
The regularized average  of $\hat{Z}$ over $\tau$ has been carried out in \cite{Dixon:1990pc}. We discuss averaging over $t$, related to it by triality, in Appendix \ref{app:n=2}. Using this result we conjecture 
\bea
T^\tau_p(\hat{Z}(\tau))={3\over \pi}\ln(p/p_0)-{3\over \pi}\ln(t_2 |\eta(t)|^4)-{3\over \pi}\ln(b_2 |\eta(b)|^4)+f(\tau)+O(1/p)
\eea
for some unknown $f(\tau)$. 
Since both  \eqref{bulkboundaryn2} and \eqref{triality} hold for any finite prime $p$, to preserve triality we must conclude $f(\tau)=-{3\over \pi}\ln(\tau_2 |\eta(\tau)|^4)$ and 
\bea
\nonumber
\overline{Z}={3\over \pi}\left.{\ln(p/p_0) - \ln(\tau_2 |\eta(\tau)|^4)-\ln(t_2 |\eta(t)|^4)- \ln(b_2 |\eta(b)|^4) \over \tau_2 |\eta(\tau)|^4}\right|_{t=t_0,\, b=b_0}+ O(1/p),
\eea
where $p_0$ is some constant. 

\subsection{Extensions and generalizations}
\label{sec:ext}
The equality between averaging over length-$n$ $\Z_p\times \Z_p$ codes (with $n\geq 3$) and ``Poincar\'e series,'' which can be understood as sums over handlebody topologies, begs a deeper understanding. First, we expect it to hold for arbitrary genus $\genus$, 
\bea
\label{PSg}
&& {1\over \mathcal N}\sum_{\CC} W_\CC(\Psi_{(c_1,\dots,c_\genus)}(\Omega)) \propto \sum_{{\g}\in Sp(2\genus,\Z)} \Psi_{(0,\dots,0)}({\g}\, \Omega), \\
&&W_\CC(\{X_{(c_1,\dots,c_\genus)}\})=\sum_{c_1,\dots,c_\genus\in \CC} X_{(c_1,\dots,c_\genus)},
\eea
where $X_{c_1,\dots,c_\genus}$ are formal variables associated with the $\genus$-tuples of codewords \cite{RUNGE1996175,Henriksson:2021qkt,Henriksson:2022dnu}.
We promote them to wavefunctions $\Psi_{c_1,\dots,c_\genus}$ of Chern-Simons theory on a genus-$\genus$ Riemann surface, hence their dependence on the period matrix $\Omega$. $\Psi_{(0,\dots,0)}$, associated with the zero codeword taken $\genus$ times, is the wavefunction of the Chern-Simons theory computed on a 3d manifold $\mathcal M$ with all $a$-cycles of $\partial \mathcal M$ contractible in the interior, which is an analog of thermal AdS. 

As in section \ref{sec:Pseries}, the Poincar\'e sum in \eqref{PSg} can be reformulated as a sum over a coset $\Gamma^*\backslash Sp(2\genus,\Z)$, where $\Gamma^*$ is a congruence subgroup of the modular group $Sp(2\genus,\Z)$ leaving $\Psi_{(0,\dots,0)}(\Omega)$ invariant. Extending the result of the $\genus=1$ case, we conjecture this subgroup to be $\Gamma^{Sp(2\genus)}_0(p)\subset Sp(2\genus,\Z)$, which we define to be the group of matrices 
\bea
\left(\begin{array}{cc}
A & B\\
C & D\end{array}\right)\in Sp(2\genus,\Z),\qquad C=0\,\,  {\rm mod}\, p.
\eea
For prime $p$ the coset $\Gamma^{Sp(2\genus)}_0(p)\backslash Sp(2\genus,\Z)$ consists of
\bea
{\mathcal N}_{sp}(\genus,p)=\prod_{j=1}^\genus(p^j+1) \label{numberofgeometries}
\eea
elements (we obtained this expression by generalizing \cite{Runge,cosnier2018bps}), which matches the result  ${\mathcal N}_{Sp}(2,2)=15$ found in \cite{Henriksson:2022dml}.

In a somewhat similar fashion, the sum over codes on the left-hand-side of \eqref{PSg} can also be represented as a coset. We recall that all even self-dual codes over $\Z_k\times \Z_k$ of the kind we are considering can be understood as a set of even self-dual lattices $\Lambda_\CC$ satisfying, see \eqref{glue},
\bea
(\sqrt{k}\Z)^{2n}\subset \Lambda_\CC \subset (\Z/\sqrt{k})^{2n}.
\eea
This defines the action of $O(n,n,\Z)\subset O(n,n,\R)$ on codes. For prime $k=p$ this action is transitive.\footnote{The action of $O(n,n,\Z)$ is also transitive on all even non-zero codewords. This is sufficient to obtain the averaged enumerator polynomial $\overline{W}$
for arbitrary $n,p$ for genus 1, thus completing the mathematical proof of \eqref{main}.}  Indeed, one can bring any code to canonical form with a generator matrix of the form $(I,{\rm B})$,where $\rm B$ is antisymmetric  mod $p$ \cite{Angelinos:2022umf}, and then use $O(n,n,\Z)$ to make $\rm B$ vanish. In other words, for prime $p$ the set of all codes can be described as a coset 
\bea
\label{coset}
{O(n,n,\Z)\over \Gamma_0^{O(n,n)}(p)},
\eea
where $\Gamma_0^{O(n,n)}(p)$, the subgroup of $O(n,n,\Z)$ which leaves the code with ${\rm B}=0$ invariant, is defined to be the group of 
matrices 
\bea
\left(\begin{array}{cc}
A & B\\
C & D\end{array}\right)\in O(n,n,\Z),\qquad C=0\,\,  {\rm mod}\, p.
\eea
The coset description \eqref{coset} is a generalization to arbitrary prime $p$ of the coset construction for $p=2$ outlined in  \cite{Dymarsky:2020qom}. The size of the coset is given by \eqref{numberofcodes}.

To summarize, the equality \eqref{PSg} between the average over codes and the Poincar\'e series over topologies can be rewritten as a sum over similar cosets
\bea
\sum_{\CC\in \Gamma_0^{O(n,n)}(p)\backslash O(n,n,\Z)} W_\CC (\{\Psi_{(c_1,\dots ,c_\genus)}\}) \propto  \sum_{{\g}\in \Gamma_0^{Sp(2\genus)}(p)\backslash Sp(2\genus,\Z)} {\g}(\Psi_{(0,\dots,0)}). \label{mainidentity}
\eea
The number of terms on both sides, ${\mathcal N}(n,p)$ \eqref{numberofcodes} and ${\mathcal N}_{Sp}({\bf g},p)$ \eqref{numberofgeometries}, and the overall similarity of the cosets, can be seen as an extension of the worldsheet/target space duality of the $c=2$ case \cite{dijkgraaf1988moduli}.

We have seen in the previous section that the Poincar\'e sum for genus one can be represented in terms of a Hecke operator. In general the Hecke operator $T_k$  is defined to act on functions $f(\Lambda)$ on lattices $\Lambda$. Then $(T_k\, f)(\Lambda)$ is a sum $f(\Lambda')$ over all sublattices $\Lambda'\subset \Lambda$ of index $k$. A modular form $f(\tau)$ can be understood as a function on two-dimensional lattices generated by $1$ and $\tau$. Then $T_k$ can be written as a sum over equivalence classes of $2\times 2$ integer matrices of determinant $k$, 
\bea
\left(\begin{array}{cc}
a & b\\
c & d
\end{array}
\right)\in M_k, \quad a,b,c,d\in \Z,\quad ad-bc=k,
\eea
modulo right multiplication by any element of $SL(2,\Z)$. For prime $k=p$ this sum includes $p+1$ terms and $T_p$ is given by \eqref{Hecke}. The equivalence of $\Gamma_0(p)\backslash SL(2,\Z)$ and $M_p/SL(2,\Z)$, together with the relation between the Poincar\'e series and the Hecke operator representation \cite{koblitz2012introduction}, leads to equality \eqref{bulknn}, which we rewrite as 
\bea
\nonumber
{1\over (1+p^{1-n})\tau_2^{n/2}|\eta(\tau)|^{2n}}\sum_{\g \in \Gamma_0(p)\backslash SL(2,\Z)}  \hat{Z}_{\vec{0},\vec{0}}(\g\, \tau)={p^{-n/2}\over (1+p^{1-n})\tau_2^{n/2}|\eta(\tau)|^{2n}}\sum_{\g\in M_p/SL(2,\Z)} \hat{Z}(\g\, \tau).\\
\label{eqP}
\eea
Here we introduced the modular invariant partition function of primaries $\hat{Z}(\tau)=\tau_2^{n/2} \Theta(\tau)$, which is related to   $\hat{Z}_{\vec{0},\vec{0}}(\tau)=\tau_2^{n/2}\Theta_{\vec{0},\vec{0}}(\tau)$ as follows
\bea
\hat{Z}_{\vec{0},\vec{0}}(\tau)=p^{-n/2}\hat{Z}(p\, \tau).
\eea
It is tempting to speculate that an analogous representation is also possible for higher-genus Poincar\'e series, in which the Hecke operator would be defined  to act on modular forms $f(\Omega)$ of $Sp(2\genus,\Z)$. 

The left-hand-side of \eqref{mainidentity}, the sum over codes, is also very reminiscent of Hecke operators. While the standard Hecke operator includes the sum over all sublattices of index $p$, the sum over even self-dual codes can be readily rewritten as a sum over all even sublattices of $(\Z/\sqrt{p})^{2n}$ of index $p^n$.  
The Hecke form of code averaging, through a suitable generalization of \eqref{clozel},  could potentially lead to a more straightforward and general proof that in the limit that the size of the code ensemble becomes infinite, the code average computes the average over the whole Narain moduli space with the Haar measure. 
We should note that formally the same logic can be applied to the Poincar\'e series, which would naively suggest that  when $p\rightarrow \infty$, the right-hand-side of \eqref{eqP} and hence \eqref{ensembleH} would be given by an integral over the fundamental domain of $\tau$. This is not so because the corresponding modular form $\hat{Z}$ is not square-integrable on the fundamental domain. As a result, in order to apply \eqref{clozel} the integral has to be covariantly regularized, 
eventually leading to the conclusion of section \ref{sec:kinf}: in the $p\rightarrow \infty$ limit the Poincar\'e series in \eqref{ensembleH} approaches the real Eisenstein series of weight $n$.  

\section{Ensemble averaging, typicality and holography}
\label{sec:typicality}
In section \ref{sec:kinf} we saw that averaging partition functions over the ensemble of code CFTs in the $k\rightarrow \infty$ limit leads (for $n>2$) to ``$U(1)$-gravity,'' the sum over CS theories on all handlebody topologies.  In particular, the answer does not depend on the embedding of the codes, and is equal to the average of the whole Narain moduli space with the Haar measure, as was outlined in  \cite{Afkhami-Jeddi:2020ezh,Maloney:2020nni}. 
This suggests that code CFTs in the $k\rightarrow \infty$ limit, when the ensemble becomes infinitely large,  densely cover the entire Narain moduli space with the canonical measure. 
This is in agreement with an earlier observation that the averaged code theory, in the $k\rightarrow \infty$ limit, has the same spectral gap as the averaged Narain theory \cite{Dymarsky:2020pzc}. If we additionally take the large central charge limit, $c\gg 1$,  then averaging over the whole moduli space  would be well approximated by a  random Narain theory, because the ensemble of all Narain theories, as well as the ensemble of code CFTs in the $c\rightarrow \infty$ limit, are self-averaging at large $c$, namely the variance is exponentially small $e^{-O(c)}$ \cite{Dymarsky:2020pzc,Collier:2021rsn,Henriksson:2022dml}.

To support the conclusion that the $k\rightarrow \infty$ ensemble densely covers the entire moduli space, we first note that there are two code ensembles, but for large $k$ they are similar. The first ensemble, which we used in our discussions above, is the ensemble of all ${\mathcal N}=\prod_{i=0}^{n-1}(p^i+1)$ codes of length $n$ (here we assume $k=p$ is prime). The second ensemble is the ensemble of all ${\mathcal N}'=p^{n(n-1)/2}$ codes in the  canonical form, also called  the $\rm B$-form \cite{Dymarsky:2020qom}. Each code in the canonical form is parametrized by an antisymmetric matrix $\rm B$ defined mod $p$, which can be interpreted as an adjacency matrix of a graph  with edges carrying an element of $\Z_p$. Every code from the first ensemble has an equivalent code in the canonical form, in the sense of code equivalences. It is a non-trivial question to determine the number of codes equivalent to the canonical one with a given $\rm B$ (noting that certain canonical form codes are equivalent to each other).  
At the level of CFTs, code equivalence is the same as T-duality only for the most symmetric ``rigid'' embedding, when the code with the matrix  $\rm B$ is associated with the Narain lattice specified by $\gamma=I/\sqrt{p}$ and $B={\rm B}/p$.  
When $p\rightarrow \infty$ we expect averaging over both ensembles with each code entering with equal weight to be physically equivalent, which is reflected by ${\mathcal N}'/{\mathcal N}\rightarrow 1$ and by the equivalence of the resulting Gilbert-Varshamov bounds (averaged spectral gap). 

The ensemble of all codes in the canonical form, independently of the embedding, leads to the ensemble of CFTs with $\gamma\rightarrow 0$ and with the $B$-field homogeneously covering the ``T-duality cube'' $B_{ij}\sim B_{ij}+1$. We {\it conjecture} that the region  in the moduli space 
\bea
\label{domain}
\gamma\rightarrow 0,\qquad 0\leq B_{ij}<1,
\eea
with the conventional flat measure for $dB_{ij}$ on the cube, 
covers (via T-duality) the whole Narain moduli space, with the canonical $O(n,n,\R)$-invariant measure. By $\gamma\rightarrow 0$ we mean that all singular values of $\gamma$ approach zero. 
This is  analogous to the observation in section \ref{c=1c=2} that the $t_2\rightarrow 0$, $0\leq t_1<1$ region is T-dual to the entire fundamental domain of $t$ with the canonical measure. Similarly here, it is straightforward to see that starting from an arbitrary  pair $(\gamma, B)$, via T-duality one can move it into the region \eqref{domain}. A non-trivial point, which we leave for the future, is whether the Haar measure on the Narain moduli space indeed results in the homogeneous measure for $B_{ij}$. With this assumption, we would find that the ensemble of all canonical codes densely covers the entire Narain moduli space with the Haar measure. In particular, this would explain why the averaged spectral gap matches the one of the whole Narain ensemble \cite{Dymarsky:2020pzc}.

The representation of the  Narain moduli space via \eqref{domain}  provides a new easy way to obtain the original result of \cite{Afkhami-Jeddi:2020ezh,Maloney:2020nni} and \cite{Datta:2021ftn}. Starting from the conventional representation of the CFT path integral 
\eqref{Psig} with self-dual $\Lambda$ parametrized by $\gamma, B$, and by performing Poisson resummation over half of the variables, we obtain 
\bea
\label{Zrep}
Z^{\gamma, B}_{BPI}(\tau, \xi ,\bar \xi)&=& {{\rm det}\, (\gamma)\over \tau_2^{n/2}|\eta(\tau)|^{2n}} \sum_{\vec{n},\vec{m} \in \Z^n} e^{-{\pi \over \tau_2}|\vec{v}|^2 -2\pi i\,  m^T B n -{2\pi \over \sqrt{2}\tau_2}(\xi \cdot v^*-\bar \xi \cdot v)+{\pi \over \tau_2}\xi \bar\xi },\\
\vec{v}&=&\gamma(\vec{n}\tau +\vec{m}). \nonumber
\eea
Now we are ready to average $Z_{\gamma, B}$ over the region \eqref{domain}.
Integration over $B_{ij}$ forces the vectors $\vec{n},\vec{m}$ to be collinear. 
We thus can parametrize $\vec{n}=c\,\vec{\ell},\, \vec{m}=d\,\vec{\ell}$ with $\vec{\ell}\in \Z^n$ and with a co-prime pair $(c,d)=1$. Using the explicit modular invariance of \eqref{Zrep},
\bea
\tau\rightarrow \tau'={a\tau+b\over c\tau+d},\quad \xi \rightarrow \xi'={\xi\over c\tau+d},\quad \bar \xi \rightarrow  \bar\xi'={\bar \xi\over c\bar \tau+d},\\  (\vec{n}\,\, \vec{m})\rightarrow (\vec{n}\, \, \vec{m})\left(
\begin{array}{cc}
a & b \\
c & d 
\end{array}\right), \nonumber
\eea
we find 
\bea
\label{onev}
\overline{Z}_{BPI}=\sum_{(c,d)=1} {{\rm det}\, (\gamma)\over \tau_2^{n/2}|\eta(\tau)|^{2n}}  \sum_{\vec{\ell}\in \Z^n}   e^{-{\pi \over \tau'_2} |\vec{v}|^2-{2\pi\over \sqrt{2}\tau_2'} (\xi' -\bar \xi') \cdot \vec{v}+{\pi \over \tau_2}\xi \bar\xi},\qquad \vec{v}=\gamma \vec{\ell}.
\eea
In the limit $\gamma \rightarrow 0$, the summation over $\ell$ can be replaced by an integration, giving
\bea
\overline{Z}_{BPI}=\sum_{(c,d)=1} {1 \over |\eta(\tau')|^{2n}}    e^{{\pi\over 2\tau'_2}(\xi'^2 +\bar \xi'^2)},
\eea
matching \eqref{flavor}.

The calculation above hints towards a possible ``holographic dual'' for an individual Narain theory, understood as a ``Poincar\'e sum'' over all co-prime pairs $(c,d)=1$, enumerating all handlebodies. The representation \eqref{Zrep}, valid for any $\gamma, B$, geometrizes the action of the modular group $SL(2,\Z)$  as an action on a lattice of vectors $(\vec{n},\vec{m})\in \Z^{2n}$. There is a trivial orbit of $SL(2,\Z)$, consisting of the origin $\vec{n}=\vec{m}=0$, with a  non-trivial action on all other elements (trivial stabilizer). Thus, only the contribution of the origin does not admit the Poincar\'e sum form, but we can make it as small as we want by using chains of T-dualities to render ${\rm det }(\gamma)$  arbitrarily small. The remaining contributions of $\Z^{2n}\backslash \vec{0}$ may be conveniently split into ``one-vector orbits'', with collinear $\vec{n}=c\,\vec{\ell},\, \vec{m}=d\,\vec{\ell}$, and ``two-vector orbits'', when $\vec{n},\vec{m}$ are not collinear.  The contribution of the one-vector orbits leads naturally to a sum over co-prime pairs $(c,d)$, as in \eqref{onev}. We can choose a representative in each orbit with $\vec{n}=0,\, \vec{m}=\vec{\ell}$, leading to a concise expression for $\vec{v}=\gamma \vec{\ell}$.  
Averaging over $\gamma$, even without assuming $\gamma \rightarrow 0$, would lead to ``$U(1)$-gravity'' -- the real Eisenstein series.\footnote{As was shown in \cite{Afkhami-Jeddi:2020ezh}, averaging over $\gamma$ with ${\rm det}\,(\gamma)$ fixed is equivalent to  replacing the sum over $\vec{\ell}\in \Z^{n}$ by an integral over $\R^n$.} 

The contribution of the two-vector orbits can also be represented as a sum over co-prime $(c,d)=1$, but here the 
choice of ``gauge'' -- the choice of a  representative in the orbit of $SL(2,\Z)$ -- is less clear. There is no obvious choice admitting an apparent bulk interpretation. 
For a typical Narain theory with large central charge, when $B$ can be considered random, the contribution of two-vector orbits will be exponentially small: for small $\gamma$ the term $-\pi |v|^2/\tau_2$ in the exponent can be neglected, while many different pairs $(n,m)$ will lead to random phases $e^{-2\pi i m^T B n}$ canceling each other. 

 To summarize, we outlined a possible holographic ``Poincar\'e sum'' representation for an individual Narain theory, which fits the picture proposed in \cite{Dymarsky:2020pzc}. A typical theory (when $c\gg 1$) will be described by $U(1)$-gravity with exponentially small corrections. There is a natural ambiguity of assigning these corrections to individual handlebodies, rooted in the ambiguity of choosing representatives among the  two-vector orbits. This, together with the need to consider a limit of T-duality transformations yielding ${\rm det}(\gamma) \rightarrow 0$, precludes a simple microscopic  local bulk interpretation. 

The resulting picture is qualitatively similar to the one in JT gravity  \cite{Blommaert:2021fob,Blommaert:2022ucs}.
An attempt to extend a holographic duality based on a sum over topologies to describe an individual Narain theory leads to potentially non-local interactions in the bulk, responsible for  ``half-wormholes.'' After averaging over all theories these interactions vanish. The crucial difference with the JT gravity case is that in our case there is also a bona fide holographic description for an individual theory, described in section \ref{sec:convhol} above, though it does not involve a sum over topologies. It would be very interesting to make an explicit link between the two holographic descriptions for a given theory, by starting from CS theory on a given handlebody and deforming it into a sum over topologies with some non-local action.

\section{Discussion}
\label{sec:Discussion}

In this paper we considered Narain CFTs on $\Sigma$, and found that they are described by pure  level-1 $(U(1)\times U(1))^n$ Chern-Simons theory on a 3d manifold $\mathcal M$ with $\partial {\mathcal M}=\Sigma$. 
The details of $\mathcal M$ do not matter; all manifolds with  $\partial {\mathcal M}=\Sigma$ lead to the same partition function because the Hilbert space of $k=1$ CS theory is one-dimensional. 
The two $U(1)^n$ gauge fields are linked at the level of large gauge transformations. The choice of large gauge transformations, or, equivalently, the choice of boundary conditions changing the representation of the unique CS wavefunction, specifies the dual Narain theory. This provides a holographic duality, with the holographic dictionary as outlined in section \ref{sec:convhol}.
Our considerations were limited to genus one $\Sigma$, but it should be straightforward to extend the duality to arbitrary genus. 

We then proceeded to consider an ensemble of Narain CFTs defined in terms  of an ensemble of codes. We considered an ensemble of all even self-dual codes of length $n$ over $\Z_p \times \Z_p$ for prime $p$, and then embedded (mapped) these codes into the $c=n$ Narain moduli space. The embedding is specified by an arbitrary ${\mathcal O}\in O(n,n,\R)$, thus any given code can be mapped to any given Narain theory. As $n$ or $p$ grows, the size of the ensemble given by \eqref{numberofcodes} grows much faster than the dimension of $O(n,n,\R)$. Hypothetically, in the  $p\rightarrow \infty$ limit, the ensemble of code theories densely covers the whole Narain moduli space with the canonical measure.  For fixed $n$ and $p$ we find that the CFT partition function averaged over this ensemble is given by the level-$p$ $(U(1)\times U(1))^n$ Chern-Simons theory summed over all classes of handlebody topologies that are distinguished by that theory.  
The main identity  \eqref{ensembleH}, valid for any fixed $n,p$ and fixed embedding, establishes an explicit relation   between averaging over the code-based ensemble and the ``Poincar\'e series'' representing the sum over topologies. Again, our explicit consideration was focused on genus one. 

One of the questions our construction answers is why the ``U(1)-gravity'' of \cite{Afkhami-Jeddi:2020ezh,Maloney:2020nni}, though suggestive, has no well-defined microscopic bulk description. In section \ref{sec:kinf} we found that  U(1)-gravity emerges as the $p\rightarrow \infty$ limit of our construction, hence it is an infinite  limit of a family of level-$p$ pure Chern-Simons theories, which  are all well-defined in the bulk. In our formalism the sum over bulk manifolds originates from a sum over $SL(2,\Z)$ transformations of a specific solid torus, and it is thus natural that we get a sum over just handlebodies and not other manifolds. Taking a leap to the holographic  CFTs of \cite{Hartman:2014oaa}, presumably dual to 3d quantum gravity with additional light matter, and the failure to find a dual to pure gravity due to intrinsic inconsistencies \cite{Maloney:2007ud,Witten:2007kt,Keller:2014xba,Benjamin:2019stq,Alday:2019vdr,Benjamin:2020mfz}, we can speculate that pure 3d quantum gravity might not be well defined by itself, but could emerge as an infinite limit of a family of well-defined theories.

From the mathematical point of view, our main technical result, equation \eqref{main}, deserves a better understanding. It would be interesting to extend it to higher genus \cite{future} and to disconnected manifolds. More generally, rewriting this equation in terms of sums over cosets or in terms of Hecke operators as was done in section \ref{sec:ext} hints at a deeper mathematical structure.  Beyond the $\Z_k\times \Z_k$ codes considered in this paper, the general code construction of \cite{Angelinos:2022umf} described in section \ref{sec:gencodes} opens up possibilities for considering other types of code ensembles. Consideration of a variety of ensemble types could help answer a crucial  question:  when is an ensemble holographic, in the sense of admitting a bulk description in terms of a sum over geometries.    

When the central charge is large $c\gg 1$,  ensembles of code CFTs or the ensemble of all Narain theories are self-averaging: a random theory faithfully captures the ensemble average up to exponentially small (in $c$) corrections. This suggests that individual theories, at least the sufficiently typical ones, should admit a bulk description in terms of a sum over topologies. We outline such a description in section \ref{sec:typicality}, but notice that it suffers from ambiguities and possibly non-local interactions in the bulk. It would be very interesting to explicitly relate this bulk description, which includes the sum over topologies, to the conventional holographic description in terms of level-1 CS theory on a fixed topology, discussed in section \ref{sec:convhol}.

Our work clarifies the role codes play in relation to CFTs and their holographic duals. We saw that all possible ``words'' label all possible wavefunctions in the bulk. We also saw that an ensemble of codes plays a crucial role in holography, although the reason why remains obscure. We emphasize that this is only one aspect of a more comprehensive story. We recall that the theory dual to the $c=1$ compact scalar,  the ``AB'' Chern-Simons theory, also emerges as a low energy limit of a 2+1 dimensional system describing Kitaev's toric code \cite{nayak2008non}. Is there a relation between the codes of this paper and the quantum codes underlying the ``AB'' theory? A first step connecting these two pictures was taken in \cite{Buican:2021uyp} for the $\Z_2\times \Z_2$ case, where classical additive codes can be understood as quantum codes. More progress followed recently, relating quantum codes (connected to  $\Z_p\times \Z_p$ classical codes in our nomenclature) to CFTs and Chern-Simons theories \cite{Kawabata:2022jxt,alam2023narain,Kawabata:2023iss}, but a complete picture is yet to emerge.   

The codes considered in this work are of additive type; consequently the corresponding CFTs are Abelian. There is a natural generalization of our story to non-Abelian codes, WZW theories and dual non-Abelian Chern-Simons theory \cite{nonabelian}. Going in the direction of gradually generalizing the type of CFTs under consideration, one hopes to eventually arrive at codes associated with the conventional ``Virasoro'' CFTs, dual to quantum gravity. We can only speculate that at this point a direct link may emerge between the code structure on the CFT side and the holographic codes responsible for the locality in the bulk \cite{Almheiri:2014lwa,Pastawski:2015qua}.

\acknowledgments
We thank Ahmed Barbar, Nathan Benjamin, Debarghya Chakraborty, Mathew Dodelson, Daniel Jafferis, Johan Henriksson, Brian McPeak, Adam Schwimmer and Edward Witten for discussions. 
The work of OA was supported in part  by an Israel Science Foundation (ISF) center for excellence grant (grant number 2289/18), by ISF grant no. 2159/22, by Simons Foundation grant 994296 (Simons
Collaboration on Confinement and QCD Strings), by grant no. 2018068 from the United States-Israel Binational Science Foundation (BSF), by the Minerva foundation with funding from the Federal German Ministry for Education and Research, by the German Research Foundation through a German-Israeli Project Cooperation (DIP) grant ``Holography and the Swampland'', and by a research grant from Martin Eisenstein. OA is the Samuel Sebba Professorial Chair of Pure and Applied Physics. 
A.D.~is grateful to Weizmann Institute of Science for hospitality and acknowledges   sabbatical support of the Schwartz/Reisman Institute for Theoretical Physics, and support by the NSF under  grants PHY-2013812 and 2310426. This work was performed in part at Aspen Center for Physics, which is supported by National Science Foundation grant PHY-2210452.  A.S. thanks the Institute for Advanced Study for hospitality and sabbatical support, and the Simons Center for Geometry and Physics for hospitality during the 2023 Simons Summer Workshop.

\appendix
\section{The compact scalar CFT} 
\label{App:Z}
The compact scalar CFT of radius $R$ is a two-dimensional theory of a real scalar field $X$, subject to the identification $X\sim X+2\pi R$, coupled to external gauge fields. The free scalar theory has a $U(1)_L\times U(1)_R$ global symmetry, and we call the corresponding charges $Q$ and $\bar Q$.
We consider the Euclidean theory placed on a spacetime torus with modular parameter $\tau$. 
The CFT partition function with fugacities (background gauge fields) $\xi$ and $\bar \xi$ is defined as a sum over the Hilbert space of the theory on a circle
\bea \label{partfunc}
Z={\rm Tr}\, \left[ q^{L_0-1/24} {\bar q}^{\bar L_0-1/24} e^{2\pi i (\xi Q-\bar \xi \bar Q)} \right],\qquad q=e^{2\pi i \tau}.
\eea
It can be readily evaluated \cite{francesco2012conformal}
\bea
Z(\tau, \xi, \bar \xi, R)={\sum_{n,m} e^{i\pi \tau p_L^2-i\pi \bar \tau p_R^2+2\pi i (p_L \xi-p_R \bar \xi)}\over |\eta(\tau)|^2},\quad 
p_{L,R}={n\over R}\pm {m R\over 2}. 
\label{partitionfunction}
\eea
The simultaneous reflection $p_R\rightarrow -p_R$, $\bar \xi \rightarrow -\bar \xi$ is a symmetry of $Z$, which is the T-duality exchanging $R$ and $2/R$, 
\bea
\label{tduality}
Z(\tau, \xi, \bar \xi, R)=Z(\tau, \xi, -\bar \xi, 2/R).
\eea

We would like to obtain the partition function \eqref{partitionfunction} from the path integral formulation.
We parametrize the spacetime torus by a complex coordinate $z$, 
\bea
z\sim z+1,\qquad z\sim z+\tau,\qquad \tau=\tau_1+i\tau_2,
\eea
with the notation
\bea
\int d^2z=\tau_2,\qquad \int dz\wedge d\bar z=-2i\tau,
\eea
where the integrals are over the torus. 
The scalar field $X(z,\bar z)$ is periodic up to identifications,
\bea
\label{periodicity}
X(z+1)=X(z)-2\pi R n_2,\quad X(z+\tau)=X(z)+2\pi R n_1.
\eea
The sign of $n_2$ is chosen for convenience. 

One way of coupling the theory to a background gauge field is by the action
\bea
\label{S1}
S[X,A]={1\over 2\pi}\int  d^2z\, \,  |\partial_z X|^2-{i\over 2\pi R} \int dX \wedge A,
\eea
where $A$ is an external $U(1)$ gauge field coupling to a specific combination of the global symmetries and satisfying $dA=0$. Using the background field gauge freedom we can choose 
\bea
\label{xiA}
\xi={\tau_2\over i\pi R} A_{\bar z},\qquad \bar \xi=-{\tau_2\over i\pi R} A_{z}
\eea
to be constant on the torus. For a real background field $A$, $\xi$ and $\bar \xi$ are complex conjugate to each other, $\xi^*=\bar \xi$. 
The theory is free (quadratic), so the partition function can be computed straightforwardly. 
We should sum over on-shell configurations satisfying \eqref{periodicity},
\bea
\label{classX}
X=2\pi R{(n_1+n_2 \bar\tau)z-(n_1+ n_2\tau)\bar z\over 2i \tau_2},
\eea
and the small fluctuations around the classical solutions contribute a multiplicative factor, that includes the Dedekind eta-function \cite{francesco2012conformal}. The full expression for the path integral is then
\bea
\nonumber
&&Z_{\rm PI}(\tau,R,\xi,\bar \xi)=\int {\mathcal D}X\, e^{-S[X,A]}=
\\&&
\qquad\qquad {R\over \sqrt{2\tau_2}|\eta(\tau)|^2}\sum_{n_1,n_2}  e^{-{\pi R^2\over 2\tau_2}|n_1+n_2\tau|^2-{\pi R\over \tau_2}\left(\xi (n_1+n_2\bar \tau)-\bar \xi (n_1+n_2 \tau)\right)}. \label{rep}
\eea 
Under large background gauge transformations $A\rightarrow A+d\phi_A$, where $\phi_A=\pi {(n+m \bar\tau)z-(n+m\tau)\bar z\over i \tau_2}$, we have from \eqref{xiA}
\bea
\label{LGT1}
\xi\rightarrow \xi+{n+m\tau\over R},\qquad \bar\xi\rightarrow \xi+{n+m\bar\tau\over R},
\eea
and the action \eqref{S1} is  shifted by an integer multiplied by $2\pi i$. Hence, the Euclidean path integral is invariant under \eqref{LGT1}, which can be verified explicitly from \eqref{rep}. Similarly, $Z_{\rm PI}$ is invariant under modular transformations generated by the two transformations
\bea
&\tau \rightarrow \tau+1,\quad  &\xi \rightarrow \xi,\qquad  \bar \xi\rightarrow \bar \xi, \\
&\tau \rightarrow -1/\tau,\quad  &\xi \rightarrow \xi/\tau,\quad  \bar \xi\rightarrow \bar \xi/\bar\tau, \nonumber
\eea
which is just the relabeling of spacetime coordinates, amended by a dilatation (which acts trivially since the theory is conformal). 

To find the relation between the path integral \eqref{rep} and the partition function \eqref{partfunc}, we perform a Poisson resummation in \eqref{partitionfunction} over $n$, which readily yields
\bea
Z=Z_{\rm PI} e^{-{\pi \over 2\tau_2}(\xi -\bar \xi)^2}. \label{PI1Z}
\eea

Alternatively we can couple a background gauge field $B$ to a different combination of the $U(1)$ global symmetries by
\bea
\label{S2}
S'[X,B]={1 \over 2\pi}\int d^2z\, |(\partial_z +R\, B_z)X|^2.
\eea
We assume $dB=0$ and use the background gauge symmetry to parametrize the background gauge field by 
\bea
\label{xiB}
\xi={\tau_2 \over i\pi }{R\over 2} B_{\bar z},\qquad \bar \xi={\tau_2 \over i\pi }{R\over 2} B_{z}.
\eea
In this case for real $B$ we have $\xi^*=-\bar \xi$, and large gauge transformations take $B\rightarrow B+d\phi_B$, where
$\phi_B=\pi {(p+q \bar\tau)z-(p+q\tau)\bar z\over i \tau_2}$,
and act as
\bea
\label{LGT2}
\xi\rightarrow \xi +{(p+q\tau)R\over 2},\qquad
\bar \xi\rightarrow \bar \xi -{(p+q\bar\tau)R\over 2}.
\eea
Clearly, the path integral 
\bea
\nonumber
&&Z'_{\rm PI}(\tau,R,\xi,\bar \xi)=\int {\mathcal D}X\, e^{-S'[X,B]}=\\
&&\qquad \qquad {R\over \sqrt{2\tau_2}|\eta(\tau)|^2}\sum_{n_1,n_2}  e^{-{\pi R^2\over 2\tau_2}|n_1+n_2\tau|^2-{\pi R\over \tau_2}\left(\xi (n_1+n_2\bar \tau)-\bar \xi (n_1+n_2 \tau)\right)+{2\pi \xi \bar \xi\over \tau_2}}
\eea 
is invariant under large gauge transformations \eqref{LGT2}, and is also modular invariant. A comparison with $Z_{\rm PI}$ yields $Z_{\rm PI}=Z_{\rm PI'}\,e^{{2\pi \over \tau_2}\xi \bar \xi}$, and 
\bea
Z=Z_{\rm PI'} e^{-{\pi \over 2\tau_2}(\xi + \bar \xi)^2}, \label{PI2Z}
\eea
in agreement with Appendix A of \cite{Kraus_2007}.

The two path integrals above are particular sections $\xi^*=\pm \bar \xi$ of a more general theory coupled to two gauge fields, $A$ and $B$, combined into one complex combination 
\bea \label{complexA}
S={1\over 2\pi} \int d^2 z {|\partial X|^2}-{i\over 2\pi} \int dX\wedge {\mathcal A} +{\kappa\over \pi}\int d^2z\, {\mathcal A^2},\\
{\mathcal A}={A\over R}+ i * B{R\over 2},\qquad {\mathcal A^2}\equiv {\mathcal A}_z  {\mathcal A}_{\bar z}.  \nonumber
\eea
Taking $\kappa=0$ or $2$ gives complexifications of the two path integrals with the actions \eqref{S1} and \eqref{S2} above. As follows from \eqref{tduality} and (\ref{PI1Z}),(\ref{PI2Z}) these two values are  T-dual to each other 
\bea
Z_{\rm PI}(\tau,\xi, \bar \xi,R)=Z'_{\rm PI}(\tau,\xi, -\bar \xi,2/R),
\eea
and each is invariant under one group of large gauge transformations, \eqref{LGT1} or \eqref{LGT2}. 
The ``symmetric'' choice
$\kappa=1$ corresponds to the {\it bulk path integral} discussed in the bulk of the paper
\bea
Z_{BPI}(\tau,\xi, \bar \xi)=Z(\tau,\xi, \bar \xi)\, e^{{\pi \over 2\tau_2}(\xi^2 + \bar \xi^2)}.
\eea
It is both modular-invariant and T-duality-invariant, and it changes covariantly (but is not invariant) under large gauge transformations of the form
\begin{eqnarray}
&&\xi={\tau_2\over i\pi }\left({A_{\bar z}\over R}+{B_{\bar z} R\over 2}\right),
\qquad \qquad\quad \, \, \bar \xi=-{\tau_2\over i\pi} \left({A_{z}\over R}-{B_{z} R\over 2}\right),\\
&&\xi \rightarrow \xi +{n+m\tau\over R}+{(p+q\tau)R\over 2},\qquad \bar\xi\rightarrow \xi+{n+m\bar\tau\over R}-{(p+q\bar\tau)R\over 2}.
\end{eqnarray}

\section{Chern-Simons theory: technical details}  
\label{App:CS}
In this section we provide additional details accompanying section \ref{sec:CS}. Our starting point is the $U(1)$ gauge field $A$ living on a three-manifold $\mathcal M$ as in subsection \ref{sec:CSU1}. We focus on the case when $\partial \mathcal M$ is a two-dimensional torus, with the same notation as in Appendix \ref{App:Z} above. The bulk theory is invariant under large gauge transformations $A\rightarrow A+\omega$ in \eqref{pilgt} when $\omega$ is a canonically normalized cohomology on $\partial \mathcal M$, namely $\omega=d\phi$ where $\phi$ is a multi-valued function winding along the cycles of $\partial \mathcal M$. When $\partial \mathcal M$ is a two-dimensional torus as above, we have explicitly $\phi=2\pi {(n+m \bar\tau)z-(n+m\tau)\bar z\over  2i \tau_2}$, from where \eqref{LGT} follows. Taking $A_{\bar z}=0$ at the boundary for simplicity, two consecutive large gauge transformations $\omega$ and $\omega'$ change the Chern-Simons action \eqref{CSaction} by 
\bea
-{ik\over 4\pi}\int\limits_{\partial \mathcal M }\omega \wedge \omega'=-i \pi k(nm'-n'm).
\eea
Thus the bulk theory is gauge-invariant for even $k$, while pure sign phase factors appear for odd $k$, related to the need to choose a spin structure.  

In the $U(1)\times U(1)$ case of subsection \ref{sec:CSU1U1}, there are two gauge fields $A,B$ subject to large gauge transformations $A\rightarrow A+\omega_A$, $B\rightarrow B+\omega_B$, where $\omega_{A,B}=d\phi_{A,B}$ are defined in Appendix \ref{App:Z} above.

One can imagine splitting $\mathcal M$ into two parts by a hypersurface. Imposing boundary conditions on the surface, and then integrating over them, should remove the split. This leads to the scalar product  $\langle \Psi|\Psi'\rangle $ discussed in the main text \cite{Elitzur:1989nr}, and the wave functions \eqref{psir} discussed there form an orthogonal basis,
\bea
\label{normalization}
\langle \Psi_r|\Psi_{r'}\rangle=\int {d^2\xi\over \tau_2}\, (\Psi_r(\xi))^* e^{-{k\pi\over \tau_2}|\xi|^2} \Psi_{r'}(\xi)=\sqrt{1\over 2k \tau_2}\, {\delta_{r,r'}\over |\eta(\tau)|^{2}}.
\eea
The integral here is over the torus of possible boundary conditions, defined by the large gauge transformations \eqref{LGT},
$\xi \sim \xi +n+m\tau$. 
In the case of $(U(1)\times U(1))^n$ discussed in section \ref{sec:CSgencase} above, the wavefunctions \eqref{Psig} also satisfy an orthogonality condition 
\bea
\int {d^{2n}\xi\,  d^{2n}\bar \xi  \over \tau_2^{2n}} (\Psi_c(\xi,\bar \xi))^* e^{-{\pi \over \tau_2}(|\xi|^2+|\bar \xi|^2)} \Psi_{c'}(\xi,\bar \xi)=\delta_{c,c'} {1\over\left( 2 |{\sf G}_{\Lambda}|^{1/2}\right)^{n}}{1\over \tau_2^{n}|\eta(\tau)|^{4n}}, 
\eea
where the integral is over the torus in the space of $\xi, \bar \xi$ variables defined by \eqref{lgtXI}.

To obtain the explicit form of the Wilson loop operators acting on the wavefunction in  the holomorphic representation \eqref{Wilsonpq}, we  take into  account that $A_{\bar z}$ (understood as a quantum operator) acts on ket vectors by multiplication, $A_{\bar z}|\Psi\rangle={i\pi \xi\over \tau_2}\Psi(\xi)$, and hence $A_z$ acts on bra vectors analogously $\langle \Psi|A_z=(\Psi(\xi){i\pi\over \tau_2}\xi)^*$. From here and by integrating by parts in \eqref{normalization} we find 
\bea
A_z|\Psi\rangle= {-i\over k}{\partial \over \partial \xi}\Psi(\xi),
\eea
which is used in \eqref{Wilsonpq}.
As we explained in the main text, this is in agreement with $A_z$ being canonically conjugate to $A_{\bar z}$, as  follows from the Chern-Simons equations of motion  \cite{Bos:1989wa}. 

\section{Narain $c=2$ theories}
\label{app:n=2}
The partition function $Z_{c=2}(\tau,t,b)$ of a central charge $c=2$ Narain theory depends on three modular parameters, $\tau$ and $t,b$ introduced in \eqref{tbdef}.
It can be written explicitly using the representation \eqref{Zrep}:
\bea
\label{sumc2}
Z_{c=2}(\tau,t,b)={b_2\over \tau_2 |\eta(\tau)|^4} \sum_{\vec{n},\vec{m}\in \Z^2}  e^{-{\pi \over \tau_2}|\gamma(\vec n\tau+\vec m)|^2-2\pi i\, b_1 \vec n \wedge \vec m}.
\eea
The moduli space of $c=2$ Narain theories is a product of two fundamental domains of $t$ and $b$, with the canonical $SL(2,\Z)$-invariant measure, modulo $\Z_2$ exchange symmetry. It is convenient to introduce 
\bea
\Theta_{c=2}(\tau,t,b)=|\eta(\tau)|^4 Z_{c=2}(\tau,t,b),
\eea
which is modular invariant under $t,b$ and is a weight 2 modular form with respect to $\tau$. 
The modular invariant combination $\hat{Z}=\tau_2 \Theta_{c=2}$ is the partition function of primaries;  it exhibits  triality -- full permutation symmetry under its arguments $\tau,t,b$  \cite{Benjamin:2021ygh} -- which is not manifest in the representation \eqref{sumc2}.

\subsection{All even self-dual $n=2$ codes over $\Z_p\times \Z_p$}
\label{sec:n=2codes}
There are $2(p+1)$ even self-dual codes over $\Z_p\times \Z_p$ with prime $p$, which can be split into 2 families. The first $p$ codes are generated through
\bea
\CC \ni c=(\vec{a},\vec{b})=G^T q,\quad q\in \Z_p^2,
\eea
by the following matrix
\bea
G=(I,{\rm B}^T),\qquad {\rm B}=r\left(\begin{array}{cc}
0 & 1\\
-1 & 0\end{array}\right)\, {\rm mod}\, p,\qquad 0\leq r<p.
\eea
One more code is generated by the matrix $G=(0,I)$. Another $p+1$ codes are obtained from the previous ones by exchanging $a_2$ and $b_2$. 

\subsection{Hecke operators and triality}
The $2(p+1)$ codes described above, once promoted to code CFTs, can be described as two families of $p+1$  theories, specified by  modular parameters
$t={r+t_0\over p}$ and $t=p\,t_0$ with fixed $b=b_0$, and 
$b={r+b_0\over p}$ and $b=p\,b_0$ with fixed $t=t_0$, where $0\leq r <p$. 
The sum over the latter (fixed $t=t_0$) series is easy to reformulate using the representation of the partition function \eqref{sumc2}. We start with 
\bea
\label{1}
{1\over p} \sum_{k=0}^{p-1}\Theta_{c=2}\left(\tau, t_0,{b_0+k\over p}\right)
\eea
and immediately conclude from \eqref{sumc2}  that the role of the sum over $k$ is to impose $n\wedge m=0\, {\rm mod}\, p$. 
This constraint means that the vectors $n\, {\rm mod}\, p$ and $m\, {\rm mod}\, p$, understood as vectors in $\Z_p^2$, are collinear. Thus, when $p$ is prime, we can write 
\bea
\label{coll}
m=r\, n+ p \tilde{m},\quad \tilde{m}\in \Z^2,
\eea
for some integer $0\leq r <p$, unless $n=0\, {\rm mod}\, p$, in which case
$n=p\, \tilde{n}$ for arbitrary $\tilde{n},m\in \Z^2$.
First we consider the latter case, and find 
\bea
\label{sec}
{b_2\over p\, \tau_2} \sum_{\tilde{n}, {m}\in \Z^2}  e^{-{\pi \over \tau_2}|p^{-1/2}\gamma(p \tilde{n}\tau+m)|^2-2\pi i\, {b_1\over p} p\, \tilde{n} \wedge m}=\Theta_{c=2}(p\, \tau,t,b). 
\label{3}
\eea
Next, we consider the case  \eqref{coll},
\bea
\sum_{r=0}^{p-1} {b_2\over p\, \tau_2} \left( \sum_{n, \tilde{m}\in \Z^2}  e^{-{\pi \over \tau_2}|p^{-1/2}\gamma(n\tau+rn+p\tilde{m})|^2-2\pi i\, {b_1\over p} p\, n \wedge \tilde{m}} -\sum_{\tilde{n}, \tilde{m}\in \Z^2} 
e^{-{\pi \over \tau_2}|p^{1/2}\gamma(\tilde{n}\tau+rn)|^2-2\pi i\, {b_1\over p} p^2\,\tilde{ n} \wedge \tilde{m}}
\right), \nonumber
\eea
where we explicitly subtracted the terms when both  $n,m=0\, {\rm mod}\, p$, which were already covered in \eqref{sec}. We can easily recognize these terms to be 
\bea
\label{4}
\sum_{r=0}^{p-1} {1\over p^2}\Theta_{c=2}\left({\tau+r\over p},t,b\right)-{1\over p} \Theta_{c=2}(\tau,t,p\, b).
\eea
Combining (\ref{1}),(\ref{3}),(\ref{4}) we find, in terms of the Hecke operator \eqref{Hecke}, 
\bea
\label{triext}
{1\over p} T_p^\tau\,  \Theta_{c=2}=T_p^b\, \Theta_{c=2}.
\eea
We label each Hecke operator by the variable it acts on; since $\Theta_{c=2}$ is a modular form of weight $2$ with respect to $\tau$ and weight $0$ with respect to $b$, $T_p^\tau$ and $T_p^b$ act differently.  The left-hand-side of \eqref{triext} is manifestly invariant under the exchange of $t$ and $b$, thus $  T_p^\tau\,  \Theta_{c=2}=p\, T_p^t\, \Theta_{c=2}=p\, T_p^b\,  \Theta_{c=2}$. This relation implies that the Hecke operators for $\tau$, $b$, and $t$ act on $\hat{Z}$ in a triality-symmetric way
\bea
T_p^\tau\,\hat{Z}=T_p^t\,\hat{Z}=T_p^b\,\hat{Z}.
\eea
This identity also follows directly from the representation in equation (3.34) of \cite{Benjamin:2021ygh}, which makes the triality explicit, if one takes into account that the Eisenstein series $E_{k}$ and the Maas cusp forms are eigenfunctions of $T_p$, 
and $T_p$ acts on the pseudo-modular form $E_2$ by shifting it by a constant, 
$T_p E_2=(p+1)E_2+{\rm const}$.

\subsection{Averaging over the moduli space}
The average of $Z_{c=2}(\tau,t,b)$ over the moduli space was considered in \cite{Dixon:1990pc}, where the integral 
over the fundamental domain of $\tau$ was regularized and evaluated to be 
\bea
\label{tauav}
\langle \hat{Z}\rangle_\tau={3\over \pi}\int{d^2 \tau\over \tau_2^2} (\tau_2 \Theta_{c=2})={3\over \pi}\left(\ln({N_\tau\over N_0}) -\ln(t_2|\eta(t)|^4)-\ln (b_2|\eta(b)|^4)\right),
\eea
where $N_\tau\rightarrow \infty$ is a regulator and $N_0$ is some constant. Here the integral over $\tau$ is over the ``keyhole'' fundamental domain, which has  volume $\pi/3$. 

To compare with the code ensemble in the $p\rightarrow \infty$ limit, we are interested in a different average, over the fundamental domains of $t$ or $b$, 
\bea
\langle Z_{c=2}\rangle_t={3\over \pi}\int{d^2 t \over t_2^2}  Z_{c=2}(\tau,t,b).
\eea
It is in principle related to \eqref{tauav} by triality. Since the latter is not manifest in \eqref{sumc2}, we perform this calculation below.  Many of the technical steps will mirror similar steps in 
 \cite{Dixon:1990pc},
in particular the splitting of the sum  in \eqref{sumc2} into three different contributions coming from the the origin $\vec{n}=\vec{m}=0$, the ``single-vector''  points $\vec{n}\parallel \vec{m}$, and the ``two-vector'' points $\vec{n}\nparallel \vec{m}$.  

\subsection*{The origin}
The contribution of the origin is simply 
\bea
{b_2\over \tau_2 |\eta(\tau)|^4}. \label{origin}
\eea
It is $t$-independent. 
Obviously, it remains the same after averaging over $t$. 

\subsection*{Contribution of the single vector orbits}
The starting point is to parametrize collinear $\vec{n}=c\,\vec{\ell}$ and $\vec{m}=d\, \vec{\ell}$ using a co-prime pair $(c,d)=1$ and an arbitrary non-zero vector $\vec{\ell}\in \Z^2$. The resulting sum is
\bea
\label{cdsum1}
\sum_{(c,d)=1} {b_2\over \tau_2 |\eta(\tau)|^4} \sum_{\vec{\ell}\in \Z^2} e^{-{\pi \over \tau'_2}|\gamma \ell|^2},\qquad \tau_2'(c,d)={\tau_2\over |c\tau+d|^2}.
\eea
Next, we parametrize a non-zero $\vec{\ell}=(\tilde{d},\tilde{c})k$ with co-prime $(\tilde{c},\tilde{d})=1$ and an arbitrary non-zero integer $k$ to find 
\bea
 \sum_{\vec{\ell}\in \Z^2} e^{-{\pi \over \tau'_2}|\gamma \ell|^2} =\sum_{(\tilde{c},\tilde{d})=1} \sum_{k\neq 0} e^{-{\pi b_2 k^2 \over \tau'_2 t_2}|\tilde{c}t+\tilde{d}|^2} \label{sv}.
\eea
Here we readily recognize $t_2'(\tilde{c},\tilde{d})=t_2/|\tilde{c}t+\tilde{d}|^2$ as being generated by a modular transformation of $t$. 
Though originally $t$ belonged to the fundamental ``keyhole domain'', the sum over co-prime pairs $(\tilde{c},\tilde{d})=1$ extends the range of $t'$  to the entire strip $|t_1|\leq 1/2$, $t_2\geq 0$. Averaging over $t$  thus gives
\bea
\left\langle  \sum_{\vec{\ell}\in \Z^2} e^{-{\pi \over \tau'_2}|\gamma \ell|^2}\right\rangle_t= {3\over \pi}\int_{-1/2}^{1/2} dt_1' \int_0^\infty {dt_2'\over (t_2')^2} \sum_{k\neq 0} e^{-{\pi b_2\over \tau_2' t_2'}k^2}={\tau_2'\over b_2}. 
\eea
Going back to \eqref{cdsum1}, we find the single-vector contribution, averaged over the fundamental domain of $t$, to be 
\bea
\label{divergentsum}
\sum_{(c,d)=1} {1\over |\eta(\tau)|^4|c\tau+d|^2}.
\eea
Of course this is merely a formal expression as it is divergent. Following \cite{Dixon:1990pc} we can regularize it by multiplying  \eqref{sv} by 
$(1-e^{-N_t/t_2'})$ where $N_t\rightarrow \infty$ is a regulator. As a result we have instead of \eqref{divergentsum}
\bea
\label{PW}
{3\over \pi^2 |\eta(\tau)|^4}\sum_{(c,d)=1}\sum_{k\neq 0} &&\left( {1\over k^2|c\tau + d|^2}- {1\over k^2|c\tau + d|^2+{N_t\tau_2\over \pi b_2}} \right)\\
&&={3\over \pi \tau_2|\eta(\tau)|^4}\left(-\ln(\tau_2|\eta(\tau)|^4)-\ln (b_2)+2\gamma+\ln({N_t\over 4\pi})  \right). \nonumber
\eea
It is interesting to note that the finite part in \eqref{PW}, which is essentially the Eisenstein series \eqref{divergentsum} regularized with help of a  Pauli-Villars-like approach, matches with the one obtained from the Kronecker limit formula, 
\bea
\sum_{(c,d)\neq (0,0)} {\tau_2^{s}\over |c\tau+d|^{2s}}={\pi\over s-1}+2\pi(\gamma-\ln (2))-\pi \ln(\tau_2|\eta(\tau)|^4)+o(s-1),
\eea
which is akin to dimensional regularization.

\subsection*{Contribution of the two vector orbits}
Our final step is the two-vector contribution with non-collinear $\vec{n}$ and $\vec{m}$. 
We can start with the same parametrization as above, $\vec{n}=(\tilde{d},\tilde{c})k$ with co-prime $(\tilde{c},\tilde{d})=1$ and nonzero $k$. Then by applying a (non-unique) $SL(2,\Z)$ transformation parametrized by $(\tilde{c},\tilde{d})$ acting on the vectors $\vec{n}$ and $\vec{m}$ we can bring the first vector to the form $\vec{n}=(k,0)$:
\bea
\sum_{(\tilde{c},\tilde{d})=1} \sum_{k\neq 0}\sum_{\vec{m}} e^{-{\pi \over \tau_2} |\gamma' (\vec{n}\tau+\vec{m})|^2-2\pi i b_1 n \wedge m}.
\eea
Here the matrix $\gamma'$ is defined the same way as in \eqref{tbdef}, but with $t$ transformed by an $SL(2,\Z)$ matrix parametrized by $(\tilde{c},\tilde{d})$. As in the previous subsection, the sum over  $(\tilde{c},\tilde{d})$ extends the domain of $t$ from the ``keyhole'' region to the strip $|t_1|\leq 1/2$, $t_2>0$. 
Let us now write $\vec{m}=(d',c')$. Then the two-vector contribution averaged over the fundamental region of $t$ is 
\bea
 {3 \over \pi}\int_{-1/2}^{1/2}dt'_1 \int_0^\infty {dt'_2\over (t'_2)^2} {b_2\over \tau_2 |\eta(\tau)|^4}  \sum_{k\neq 0} \sum_{c'\neq 0,\,  d'} e^{-{\pi b_2\over \tau_2 t'_2}(|k \tau+c' t'_1+ d'|^2+(c' t'_2)^2)-2\pi i\, b_1 k c'}.
\eea
In the sum above we must keep $c'\neq 0$ lest the vectors $\vec{n},\vec{m}$ become collinear. The sum over $d'$ is not restricted. We can represent it as $d'=c' r+d''$
where $r\in \Z$ and $d''$ is an integer between $0$ and $c'-1$. We can now combine 
\bea
c' t'_1+d'=c'(t'_1+r)+d'',
\eea
and the sum over $r$ plus the integral over the strip $|t_1|\leq 1/2$, $t_2>0$  become an integral over the whole upper half-plane of $t'$. The dependence on $d''$ disappears and the sum over $d''$ simply gives a factor of $|c'|$,
\bea
{3 \over \pi} \int_0^\infty {dt'_2\over (t'_2)^2}{ \sqrt{b_2 t'_2} \over \sqrt{\tau_2} |\eta(\tau)|^4} \sum_{k\neq 0} \sum_{c'\neq 0} e^{-{\pi b_2\over \tau_2 t'_2}((k \tau_2)^2+(c' t'_2)^2)-2\pi i\, b_1 k c'}.
\eea
At this point we can integrate over $t_2$, 
\bea
{3 \over \pi } {1\over \tau_2 |\eta(\tau)|^4} \sum_{k\neq 0} \sum_{c'\neq 0} e^{-2\pi b_2 |k c'|-2\pi i\, b_1 k c'}{1\over |k|}.
\eea
There are four ``branches'' with positive and negative $k$ and $c'$, which we combine into a sum of the form 
\bea
{6 \over \pi  \tau_2|\eta(\tau)|^4} \sum_{k> 0} \sum_{c' > 0} {e^{2\pi i\, b\, k c'} +e^{2\pi i\, \bar b\, k c'} \over k}.
\eea
Now we introduce $q_b=e^{2\pi i b}$ and sum over $k$ using 
\bea
\sum_{c',k>0} {q_b^{kc'}\over k}+{\rm c.c}=-\ln \prod_{c'=1}^\infty(1-q_b^{c'})+{\rm c.c}={i\pi \over 12}b-\ln (\eta(b))+{\rm c.c}
\eea
Finally, we find for the two-vector contribution, averaged over $t$, 
\bea
\label{2vecanswer}
{1\over \tau_2|\eta(\tau)|^4}\left(-b_2-{3\over \pi}\ln|\eta(b)|^4\right).
\eea

Combining everything together, we find that the first term in \eqref{2vecanswer} exactly cancels the ``origin'' contribution \eqref{origin}.  Hence, $Z_{c=2}$ averaged over the modular parameter $t$ and covariantly regularized is 
\bea
\langle Z_{c=2}(\tau,t,b) \rangle_{t}=
{3\over \pi \tau_2|\eta(\tau)|^4}\left(\ln({N_t\over N_0}) -\ln(\tau_2|\eta(\tau)|^4)-\ln (b_2|\eta(b)|^4)\right),
\eea
where $N_t\rightarrow \infty$ is a regulator and $N_0=4\pi e^{-2\gamma}$. 
The final expression is in full agreement with \eqref{tauav}. 

\subsection{Large-$p$ limit}
To evaluate the large-$p$ limit of $T_p \hat{Z}$ we first approximate it by the regularized integral over the fundamental domain 
\bea
\label{TPZ}
T^\tau_p(\hat{Z})\approx {3\over \pi}\int_{\mathcal F} {d^2\tau'\over (\tau'_2)^2} \hat{Z}(\tau')\left(1-e^{-N/\tau'_2}\right).
\eea
The value of $N$ can be fixed as follows. Modular transformations mapping $(\tau+k)/p$ back to the fundamental keywhole domain $\mathcal F$ will be more dense in the region of small $\tau'_2$, with only one point  reaching the maximal value of $\tau'_2=p/\tau_2$. Thus we can take $N\propto p/\tau_2$, leading to, c.f.~\eqref{tauav},
\bea
{3\over \pi}\left(\ln (p/p_0)-\ln(\tau_2)-\ln(t_2|\eta(t)|^4)-\ln (b_2|\eta(b)|^4) \right)+\dots
\eea
This expression is not modular invariant with respect to $\tau$, although the left-hand side of \eqref{TPZ} is, which suggests there might be additional $\tau$-dependent  finite terms. We therefore conjecture 
\bea
T^\tau_p(\hat{Z}(\tau))={3\over \pi}\ln(p/p_0)-{3\over \pi}\ln(t_2 |\eta(t)|^4)-{3\over \pi}\ln(b_2 |\eta(b)|^4)+f(\tau)+O(1/p),
\eea
where the crucial assumption is that $f(\tau)$ does not depend on $t$ and $b$. The rest follows from the extension of  triality \eqref{triext}, 
\bea
g(\tau)=-{3\over \pi}\ln(\tau_2 |\eta(\tau)|^4).
\eea

\bibliographystyle{JHEP}
\bibliography{codes}

\end{document}